\documentclass{aa}  
\usepackage{natbib,twoopt}
\usepackage[breaklinks=true]{hyperref} 
\bibpunct{(}{)}{;}{a}{}{,}             
\makeatletter
  \newcommandtwoopt{\citeads}[3][][]{\href{http://adsabs.harvard.edu/abs/#3}%
    {\def\hyper@linkstart##1##2{}%
     \let\hyper@linkend\@empty\citealp[#1][#2]{#3}}}
  \newcommandtwoopt{\citepads}[3][][]{\href{http://adsabs.harvard.edu/abs/#3}%
    {\def\hyper@linkstart##1##2{}%
     \let\hyper@linkend\@empty\citep[#1][#2]{#3}}}
  \newcommandtwoopt{\citetads}[3][][]{\href{http://adsabs.harvard.edu/abs/#3}%
    {\def\hyper@linkstart##1##2{}%
     \let\hyper@linkend\@empty\citet[#1][#2]{#3}}}
  \newcommandtwoopt{\citeyearads}[3][][]%
    {\href{http://adsabs.harvard.edu/abs/#3}
    {\def\hyper@linkstart##1##2{}%
     \let\hyper@linkend\@empty\citeyear[#1][#2]{#3}}}
\makeatother

\usepackage{graphicx}
\usepackage{txfonts}
\usepackage{subfigure}
\usepackage{longtable}
\usepackage{tabularx, blindtext}
\usepackage{rotating}
\usepackage{pdflscape}

\begin{document} 
\title{Confirmation and physical characterization of the new bulge globular cluster Patchick 99 from the VVV and Gaia surveys}
   \author{E.R. Garro \inst{1}
          \and
          D. Minniti\inst{1,2}
         M. Gómez\inst{1}
          \and
           J. Alonso-García\inst{3,4}
           \and
          T.~Palma\inst{5}
          \and
          L. C. Smith\inst{6}
          \and
          V. Ripepi \inst{7}
          }
   \institute{Departamento de Ciencias Físicas, Facultad de Ciencias Exactas, Universidad Andres Bello, Fernández Concha 700, Las Condes, Santiago, Chile
   \and
 Vatican Observatory, Vatican City State, V-00120, Italy
 \and
 Centro de Astronomía (CITEVA), Universidad de Antofagasta, Av. Angamos 601, Antofagasta, Chile
 \and
 Millennium Institute of Astrophysics, Santiago, Chile
  \and
Observatorio Astron{\'{o}}mico, Universidad Nacional de C{\'{o}}rdoba, Laprida 854, C{\'{o}}rdoba, Argentina
 \and 
 Institute of Astronomy, University of Cambridge, Madingley Rd, Cambridge CB3 0HA, UK
 \and
INAF-Osservatorio Astronomico di Capodimonte, Salita Moiariello 16, 80131, Naples, Italy
}
  \date{Received 25 August 2020; Accepted 2 March 2021}

 
  \abstract
   {
Globular clusters (GCs) are recognised as important tools to understand the formation and evolution of the Milky Way (MW) because they are the oldest objects in our Galaxy. Unfortunately, the known sample in our MW is still incomplete, especially towards the innermost regions, due to the high differential reddening, extinction, and stellar crowding. Therefore, the discovery of new GC candidates and the confirmation of their true nature are crucial for the census of the MW GC system.}
   {
   Our main goal is to confirm the physical nature of two GC candidates: Patchick 99 and TBJ 3. They are located towards the Galactic bulge. We use public data in the near-infrared (IR) passband from the VISTA Variables in the Via Láctea Survey (VVV), VVV eXtended Survey (VVVX) and the Two Micron All Sky Survey (2MASS) along with deep optical data from the Gaia Mission DR 2, in order to estimate their main astrophysical parameters, such as reddening and extinction, distance, total luminosity, mean cluster proper motions, size, metallicity and age.
   }
   {
   We investigate both candidates at different wavelengths, allowing us to discard TBJ3 as a possible GC. We use near-IR $(K_{s}\ vs.\ (J-K_{s}) )$ and optical $(G\ vs.\ (BP-RP))$ colour-magnitude diagrams (CMDs) in order to analyse Patchick 99. First, we decontaminate CMDs, following a statistical procedure, as well as selecting only stars which have similar proper motions (PMs) and are situated within $3'$ from the centre. Mean PMs are measured from Gaia DR 2 data.
Reddening and extinction are derived by adopting optical and near-IR reddening maps, and then we use them to estimate the distance modulus and the heliocentric distance. Metallicity and age are evaluated by fitting theoretical stellar isochrones.
   }
   {
Reddening and extinction values for Patchick 99  are $E(J-K_{s})=(0.12\pm 0.02)$ mag and $A_{K_s}=(0.09\pm 0.01)$ mag from the VVV data, whereas we calculate $E(BP-RP)=(0.21\pm 0.03)$ mag and $A_{G}=(0.68\pm0.08)$ mag from Gaia DR 2 data. 
We use those values and the magnitude of the RC to estimate the distance, finding a good agreement between the near-IR and optical measurements. In fact, we obtain $(m-M)_{0}=(14.02\pm 0.01)$ mag, equivalent to a distance $D=(6.4\pm 0.2)$ kpc in near-IR and $(m-M)_{0}=(14.23\pm 0.1)$ mag and so $D=(7.0\pm 0.2)$ kpc in optical. In addition,
we derive the metallicity and age for Patchick 99  using our distance and extinction values and fitting PARSEC isochrones. We find $[Fe/H]=(-0.2\pm 0.2)$ dex and $t=(10\pm 2)$ Gyr. The mean PMs for Patchick 99  are $\mu_{\alpha}=(-2.98 \pm 1.74)$ mas $yr^{-1}$ and $\mu_{\delta}=(-5.49 \pm 2.02)$ mas $yr^{-1}$, using the Gaia DR 2 data. They are consistent with the bulge kinematics. We also calculate the total luminosity of our cluster and confirm that it is a low-luminosity GC, with $M_{K_s}=(-7.0\pm 0.6)$ mag. The radius estimation is performed building the radial density profile and we find its angular radius $r_{P99}\sim 10'$. We also recognise seven RR Lyrae star members within 8.2 arcmin from the Patchick 99  centre,  but only three of them have PMs matching the mean GC PM, confirming the distance found by other methods.
}
{
 We found that TBJ 3 shows mid-IR emissions that are not present in GCs. Hence, we discard TBJ 3 as GC candidate and we focus our work on Patchick 99. We conclude that Patchick 99  is an old metal-rich GC, situated in the Galactic bulge. TBJ 3 is a background galaxy.}
   \keywords{Galaxy: bulge – Galaxy: center -- Galaxy: stellar content – Stars Clusters: globular – Infrared: stars – Surveys}

\titlerunning {Confirmation of the new bulge globular cluster Patchick 99 from the VVV and Gaia surveys}
\authorrunning {E.R. Garro et al.}
   \maketitle
\section{Introduction}
Globular Clusters (GCs) represent the most ancient stellar systems in the Milky Way (MW). For this reason, they are excellent tracers of the formation and chemodynamical evolution of the Galaxy, going from the outer halo to the innermost Galactic regions. In addition, they
survived different dynamical processes, like dynamical friction, disk and bulge shocking, evaporation that also led some ancient GCs to destruction \citep{1994AJ....108.1414W}. Today,  we distinguish different kinds of GCs, depending on their metallicities, ages and orbits. The metal-poor GCs could be the oldest population \citep{2006A&A...449.1019B, 2009A&A...507..405B}. More precisely, in the Galactic bulge, they may constitute the first stellar generation whose origin could coincide with the formation of the bulge itself or before the actual configuration of the bulge/bar component, examples are NGC~6522, NGC~6626 and HP~1 \citep{Kerber_2018, 10.1093/mnras/stz003}. On the other hand, the metal-rich GCs may be the second generation in the MW \citep{Muratov2010}. Furthermore, recent studies have demonstrated that some other luminous GCs may be actually fossil relics of accreted dwarf galaxies, such as Terzan 5 \citep{2009Natur.462..483F}, Liller 1 \citep{Ferraro2020}, Terzan 9 \citep{2019A&A...632A.103E} and also Omega Centauri \citep{1999Natur.402...55L,2008MmSAI..79..342S}. Further,  with the advent of Gaia Mission and deep surveys, it is now possible to associate individual GCs with specific merger events (e.g., \citealt{10.1093/mnras/stz171, 10.1093/mnras/stz1770, 2019A&A...630L...4M}).\\

However,  it seems that the frequency of the MW's GCs is lower than the Andromeda's GC system \citep{2010AJ....140.1043V} and the MW's open clusters.  Indeed, the total number of Galactic GCs \citep{Harris_2013} is about half those of our neighbour Andromeda \citep{Barmby_2001, 10.1093/mnras/stu771}, and the number of Galactic open clusters is vastly higher than that of GCs \citep{2010A&A...521A..74B}. There are a number of issues that make their detection a challenging task, certainly we are still missing a number of them due to high stellar extinction and stellar crowding and very low-luminosities of some of them. Nevertheless, in the last decades, the number of star cluster candidates reported in the literature has increased significantly. Indeed, recent near-infrared (IR) surveys, like the Two Micron All Sky Survey (2MASS; \citealt{2006AJ....131.1163S}) and the VISTA Variables in the Via Láctea Survey (VVV; \citealt{2010NewA...15..433M}), and mid-IR survey, like  Wide-field Infrared Survey Explorer (WISE; \citealt{2010AJ....140.1868W}) have allowed us to search for new GC candidates. In combination with near/mid-IR surveys, optical surveys are fundamental to finding new GCs, like Gaia Mission DR 2 \citep{2018A&A...616A...1G}, which provides optical photometry and proper motions (PMs). Thus, new GC candidates are being discovered, in particular in the Galactic bulge, using both the optical and near-IR images in order to help complete the census of GCs in the MW. Examples of recently discovered GCs toward to innermost Galactic regions are VVV-CL001 \citep{2011A&A...527A..81M}, VVV-CL002 and VVV-CL003 \citep{2011A&A...535A..33M}, Camargo 1102, 1103, 1104, 1105 and 11061 \citep{Camargo_2018}, Camargo 1107, 1108 and 1109 \citep{CamargoMinniti_2019}, and several tens of candidates in \cite{Minniti_2017, 2017RNAAS...1...16M, 2017AJ....153..179M}  and \cite{10.1093/mnras/stz1489}.  Even so, the nature of some new GC candidates has yet to be confirmed and main parameters have still to be estimated. \\


In this paper, we focus on two recently discovered and not yet well characterised GC candidates, Patchick99 and TBJ 3 (a.k.a. TJ 3).  We are studying for GC candidates in order to investigate the GC luminosity function in the Galactic bulge, especially at very faint luminosities. For that purpose,  Patchick 99  and TBJ 3 were selected from \cite{Bica_2018}, who compiled a list of 10,000 good star cluster candidates in the MW, as they had not been studied before. We use a combination of optical and near-IR catalogues, aiming on having a more complete information about them. 
Our main goals are to confirm their true nature as GC and estimate their main parameters  for the first time, such as reddening, distance modulus, heliocentric distance, size, total luminosity, mean cluster proper motions, metallicity and age. \\

In Section 2, the observational data are briefly described. In Section 3 we focus our attention to TBJ 3. Section 4 presents the methods used to estimate the physical parameters and the resulting values for Patchick 99.  In Section 5, a summary and conclusions are drawn.

\section{Observational Data}
The effects of dust become increasingly prominent towards the centre of our Galaxy. Thus, we have to overcome problems like high differential reddening and extinction (e.g., \citealt{Alonso_Garc_a_2017}), which introduce systematic errors in the determination of distances. Additionally, uncertainties in the extinction values affect the estimation of ages when isochrones are used. Also, the stellar density is very high in the inner regions and crowding does not always allow to resolve individual stars. Further uncertainties are due to the foreground and background field contamination that may alter the measurement of the total luminosity.
For these reasons, we use combinations of optical and near-IR data in order to substantially reduce uncertainties in the measurements and obtain robust results. \\
We analyse especially red clump (RC) and red giant branch (RGB) stars, since they are very bright in infrared and they allow us to estimate accurate values of reddening, extinction, distance and size for our GC candidates.\\

Below a short description of the observational material is provided.

\subsection{Infrared Surveys: the VVV, VVVX and 2MASS}
We use deep near-IR data from the VVV and VVVX Survey \citep{2010NewA...15..433M,2018ASSP...51...63M}, acquired with the VISTA InfraRed CAMera (VIRCAM) at the 4.1m wide-field Visible and Infrared Survey Telescope for Astronomy (VISTA; \citealt{2010Msngr.139....2E}) at ESO Paranal Observatory. VVV and VVVX data are reduced at the Cambridge Astronomical Survey Unit (CASU; \citealt{10.1117/12.551449}) and further processing and archiving is performed with the VISTA Data Flow System (VDFS; \citealt{2012A&A...548A.119C}) by the Wide-Field Astronomy Unit and made available at the VISTA Science Archive\footnote{http://horus.roe.ac.uk/vsa/}. We use preliminary data from VIRAC \citep{2018MNRAS.474.1826S} version 2, described in detail in Smith et al. (in preparation).In summary, VIRAC v2 is based on a psf fitting reduction of VVV and VVVX images using DoPhot \citep{1993PASP..105.1342S, 2012AJ....143...70A, 2018A&A...619A...4A}. Their astrometry are calibrated to the Gaia DR 2 \citep{2018A&A...616A...1G} astrometric reference system, and their photometry are calibrated against 2MASS using a globally optimised model of frame-by-frame zero points plus an illumination correction. The VVV and, in particular, the VVVX Surveys are sampling the Galactic bulge and the Southern Galactic Plane, using the $J$ (1.25 $\mu m$), $H$ (1.64 $\mu m$)), and $K_s$ (2.14 $\mu m$)) near-IR passbands, aiming to better disentangle and characterise the stellar populations located at these low latitude regions. The VISTA Telescope tile field of view is 1.501 deg$^2$, hence 196 tiles are needed to map the bulge area and 152 tiles for the disk. Adding some X and Y overlap between tiles for a smooth match, the area of our unit tile covered twice is 1.458 deg$^2$. Specifically, Patchick 99  is located in the VVV tile b251, whereas TBJ 3 is situated in the VVVX tile b462. \\

In addition, we use 2MASS data \citep{2006AJ....131.1163S, 2003yCat.2246....0C}. It is a near-IR Survey of the sky in $J$ ($1.25\ \mu m$), $H$ ($1.65\ \mu m$) and $K_s$ ($2.17\ \mu m$) passbands. We adopt it specifically to extend the RGB VVV colour-magnitude diagrams (CMDs), as these stars are saturated for $K_s<11$ mag in VVV images.  We downloaded the 2MASS data from the VizieR Online Data Catalogue \footnote{https://vizier.u-strasbg.fr/viz-bin/VizieR}, and we merge it with the VVV catalogue, including all stars located within $10'$ from the targets centres.  However, since the magnitude scale is different in the two photometric systems,  we have transformed the 2MASS photometry into the VVV magnitude scale \footnote{http://casu.ast.cam.ac.uk/surveys-projects/vista/technical/photometric-properties}.

\subsection{Optical Surveys: Gaia DR 2}
The Gaia astrometric mission maps the MW at optical wavelengths, with the aim of revealing motions, composition, formation and evolution of the Galaxy. It provides very accurate position, PMs and radial velocity measurements for sources both in our Galaxy and throughout the Local Group. We use Gaia DR 2 data \citep {2018A&A...616A...1G} for both stellar cluster candidates, downloaded from the VizieR Online Data Catalogue. Gaia~DR~2 data have been prepared by the Gaia Data Processing and Analysis Consortium (DPAC) and contain G-band magnitudes for over $1.6 \times10^9$ sources with $G < 21$ mag and broadband colours $G_{BP}$ (330-680 nm) and $G_{RP}$ covering (630-1050 nm) for $1.4\times10^9$ sources. PM components in equatorial coordinates are available for $1.3\times10^9$ sources, with an accuracy of 0.06 $mas\ yr^{-1}$, 0.2 $mas\ yr^{-1}$, and 1.2 $mas\ yr^{-1}$, for sources with $G< 15$ mag, $G\sim17$ mag, and $G\sim20$ mag, respectively. We use the parallax (\textit{plx}) values to make the first cut in such a way to exclude clear nearby foreground stars
(those with large errors might scatter to $plx<0.5$ mas).  In our analysis, we did not make any Gaia photometric or colour cuts. 

\section{TBJ 3: background galaxy}
TBJ 3 is situated at Equatorial coordinates $R.A.=17h18m14s$ and $DEC=-27d54m15s$ (J2000) and Galactic coordinates $l=357^{\circ}.9299$ and $b=5^{\circ}.0084$.  It was mentioned for the first time as a GC candidate in the list of \cite{1978Msngr..15...14T} and \cite{1980Msngr..20....6T} with the name TJ~3. Subsequently, it was spectroscopically studied by \cite{1998A&AS..131..483B}, who rename it as TBJ~3. We believe it to be the same object, although in those works the Equatorial coordinates are $R.A.(1950)=17h15m06s$ and $DEC(1950)=-27^{\circ}51'$, that correspond to $R.A.(J2000)=17h17m05.17s$ and $DEC (J2000)=-27d47m49.22s$, different from those considered in this paper. \\

\begin{figure}[h]
\centering
\includegraphics[width=4cm, height=4cm]{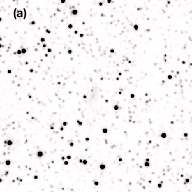} 
\includegraphics[width=4cm, height=4cm]{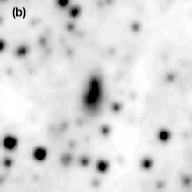} 
\includegraphics[width=4cm, height=4cm]{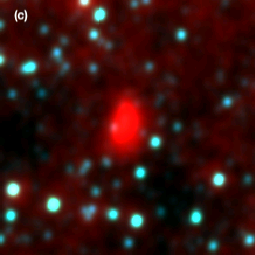} 
\includegraphics[width=4cm, height=4cm]{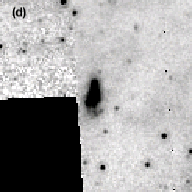} 
\caption{Finding charts of TBJ 3, adopting different Surveys: 2MASS K-band (a), ALLWISE w3 (b), ALLWISE w4 (c), Spitzer MIPS24 (d).}
\label{tbj3allwise}
\end{figure}

From Figure \ref{tbj3allwise}, it is clear that there is an object at these coordinates. Analysing it at different wavelengths and by looking especially at the WISE satellite images, we noted that TBJ 3 shows diffuse mid-IR emission, which is not present in GCs. This feature, as well as the oblate and very elliptical shape, lead us to discard TBJ 3 as a possible GC candidate, and consider this to be a background galaxy.

\section{Confirmation of Patchick 99  as a new GC}
\begin{figure*}[h]
\centering
\includegraphics[width=18cm, height=5cm]{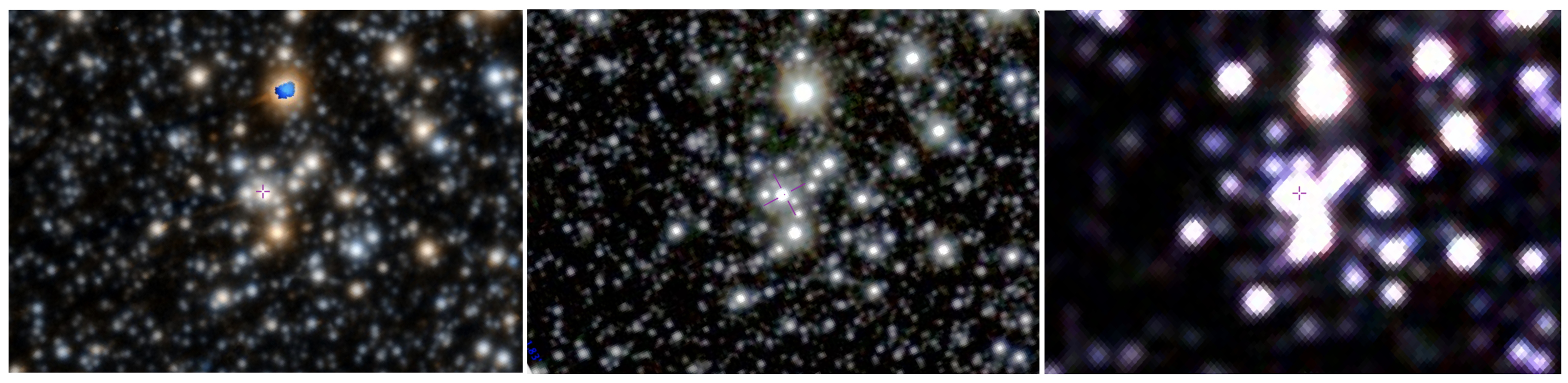} 
\caption{Finding chart of Patchick 99  showing from left to right the central $1.6'\times 1.1’$ optical colour images from PanStarrs, and near-IR colour images from the VVV and 2MASS surveys.}
\label{pos_Pat99}
\end{figure*}

\begin{figure*}[h]
\centering
\includegraphics[width=15cm, height=8cm]{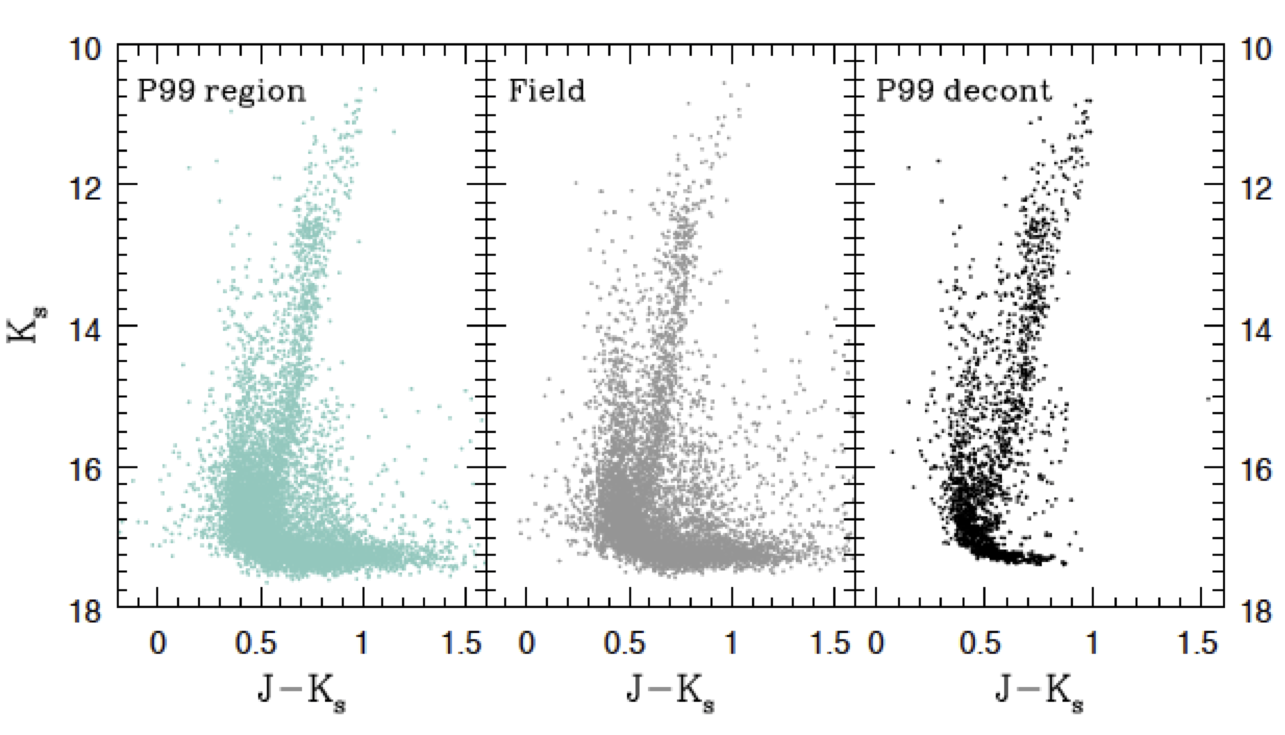} 
\caption{Observed VVV near-IR CMDs of a $3'$ radius region centred on Patchick 99 (left), of a background region with similar area (middle), and statistically decontaminated CMD (right) following the procedure described by \cite{PALMA201650,10.1093/mnras/stz1489}. We note the clear presence of the cluster RC in the decontaminated diagram. }
\label{statdec}
\end{figure*}
Patchick 99  is located in the Galactic bulge at Equatorial coordinates $R.A.=18h15m47s$ and $DEC=-29d48m46s$ (J2000) and Galactic coordinates $l=2^{\circ}.4884$ and $b=-6^{\circ}.1452$ (Figure \ref{pos_Pat99}).  We found this cluster during our VVV search \citep{Minniti_2017,2018A&A...616A..26M,2018ASSP...51...63M}, but later realized that an object at that position had been reported before as part of the deep sky hunters (DHS) survey of the DSS and 2MASS images \citep{2006A&A...447..921K,2012ASSP...29..105K},  so their discovery and nomenclature take precedence. This cluster,  which is named after the researcher who discovered it, was then listed in the compilation of \cite{Bica_2018},  but has not been studied in detail, and we therefore decided to investigate its nature and measure its physical parameters with the VVV data.
\\
\subsection{Statistical and PM decontamination procedures}
\begin{figure}[h]
\centering
 \includegraphics[width=8cm, height=8cm]{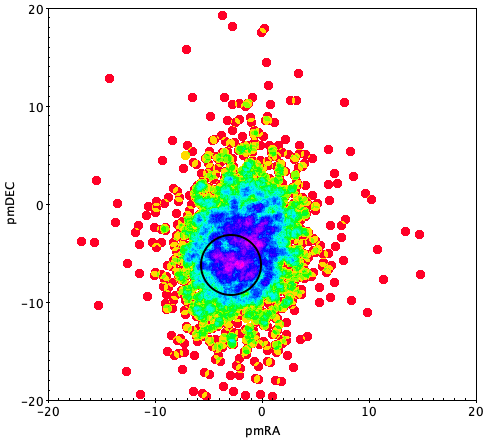} 
 \includegraphics[width=8cm, height=8cm]{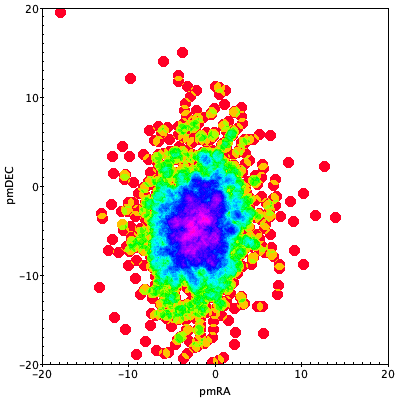} 
\caption{\textit{Top panel}: VPM diagram from Gaia~DR~2 dataset for star members located within $3'$ from the Patchick 99  centre. The black circle indicates the position of the cluster selection within $3\ mas\ yr^{-1}$.  \textit{Bottom panel}: VPM diagram for field stars located at $4.8'$ from the Patchick 99  centre.  In each panel, colour represents density, with magenta the highest, followed by blue, green and red.}
\label{pm_Pat99}
\end{figure}

\begin{figure}[h]
\centering
 \includegraphics[width=9cm, height=8cm]{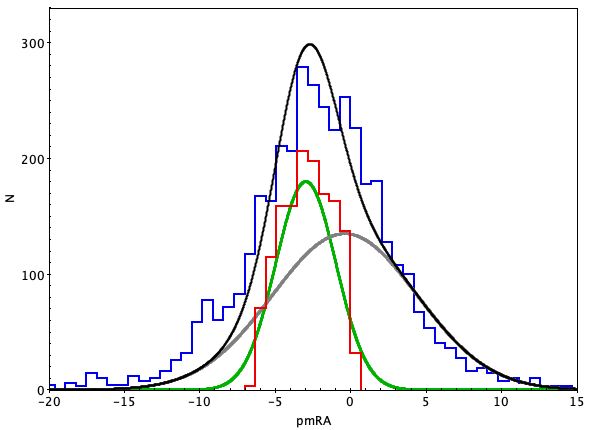} 
  \includegraphics[width=9cm, height=8cm]{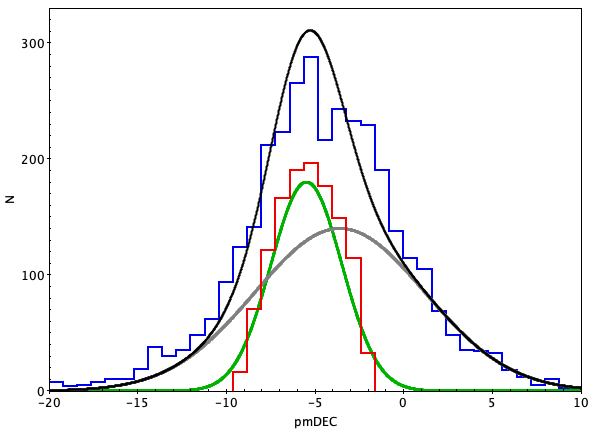} 
\caption{PM distributions in RA (top panel) and in DEC (bottom panel),  for stars located within $3'$ from the Patchick 99  centre (blue histogram) and for our cluster PM selection (red histogram).  Grey and green curves are the Gaussian distributions for the Bulge field stars and Patchick 99  stars, respectively.  The total distribution, obtained by coadding the two Gaussian distributions, is fitted as black curve. }
\label{pm_cont}
\end{figure}

Contamination by Galactic bulge and disk stars is relevant, so we have made some selections in order to decontaminate the CMDs. As mentioned in Section 2.2, we consider only stars that have $plx<0.5$ mas, in order to exclude foreground stars. 
Moreover, Figure \ref{statdec} shows the VVV near-IR CMDs of a $3'$ radius region centred on Patchick 99 compared with the CMD of a nearby background region, and with the statistically decontaminated CMD. This last CMD was constructed following the procedure described by \cite{Piatti_Bica_2012} as applied by \cite{PALMA201650,10.1093/mnras/stz1489} and \cite{Minniti_2017}. We notice clearly the presence of the cluster RGB and RC that remain in the decontaminated diagram, suggesting that a distance of $3'$ from the centre is an excellent compromise for our selection. However, there are some blue stars outside of the cluster RGB that probably belong to the foreground MW disk population that persist in this CMD. Therefore, we decided to use the PMs as an independent method to decontaminate the CMD for this cluster, since stars with similar PMs ensure belonging to the same cluster. 
In fact, we compared the vector PM (VPM) diagram for our cluster, including stars in the $3'$ radius region and for the field stars, located between $4 < r < 5$ arcmin away from the cluster centre (Figure \ref{pm_Pat99}). Visually, the main difference between the two PM distributions is that in the case of our cluster selection, this one shows two over-densities, suggesting a bimodal distribution, unlike the field stars that show a shallower homogeneous distribution when we move away from the centre of the cluster, indicating an unimodal distribution. In order to confirm our statement we have carried out the Gaussian mixture modelling (GMM) analysis \citep{Muratov2010}, which helps to give us the statistical significance of an unimodal versus a bimodal distribution. We have performed the GMM analysis with the heteroscedastic statistic for each sample.  In summary, we perform the GMM analysis for our three subsamples: the field stars located between $4<r<5$ arcmin away from the Patchick 99 centre, the position selection including all stars situated within $3'$ from the cluster centre, and our PM selection.  Firstly, we do this for field stars and position selection in order to obtain the statistical significance for the two Gaussian distributions shown in Figure \ref{pm_Pat99}. Subsequently, we run the GMM algorithm to distinguish the two Gaussian peaks shown in Figure \ref{pm_cont}. The statistical parameters are shown in Tables \ref {GMM_ra} and \ref {GMM_dec}. For a mixture of two normal distributions the means and standard deviations along with the mixing parameter (D,  separation of the means relative to their widths) are usually used, as a total of five parameters. \cite{Ashman1994} noted that $D > 2$ is required for a clean separation between the modes, while if the GMM method detects two modes, but they are not separated enough ($D < 2$), then such a split is not meaningful. \\
Applying the GMM analysis to our samples we noted that they may show a bimodal distribution, but with a low significance level.  Doing a deep analysis of the GMM resulting parameters, we can confirm that the distribution of field stars located away from the cluster centre is unimodal since the $D < 2$ and the standard deviation of the second mode is of $\sigma_{2}\approx 14$, which is not physical.  About the $3'$ radius region sample, the analysis becomes more complicated.  We believe that the two peaks have a very small separation and the method is not able to distinguish both properly, indeed comparing the parameters of the first Gaussian with the second one we can appreciate that the first dominates over the second. On the other hand, the GMM method on the PM selection indicates a bimodal distribution, since $D>2$. This may be misleading to interpret since we should have expected an unimodal distribution for the stars in a cluster. However, we have to highlight that any sample will be contaminated by foreground stars that will have different velocity (and PM) distribution. Thus,  having two peaks in the sample of PM selection is not unexpected.  In any case, this does not represent a problem since the second mode is very narrow and is not dominant over the first, meaning that our final sample has got a very small contamination. \\

\begin{figure}[h]
\centering
\includegraphics[width=4.4cm, height=5cm]{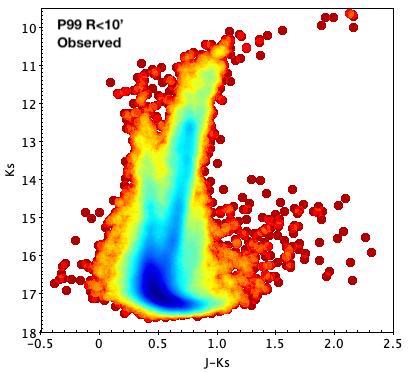}
\includegraphics[width=4.4cm, height=5cm]{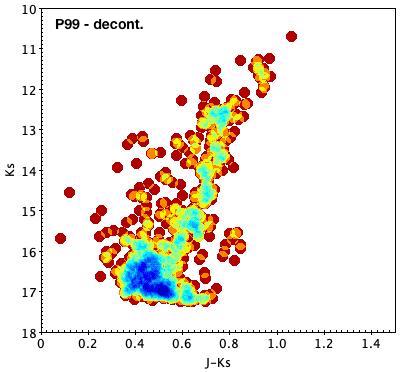} 
\includegraphics[width=4.4cm, height=5cm]{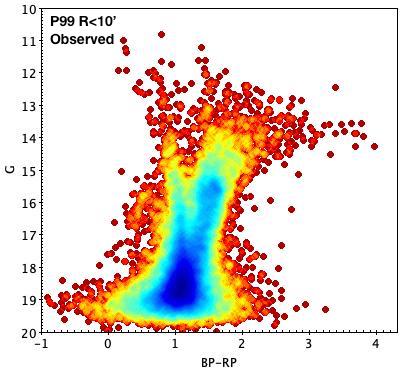}
\includegraphics[width=4.4cm, height=5cm]{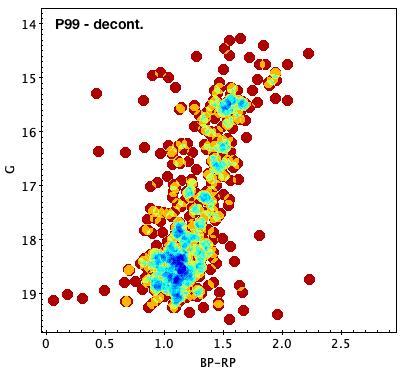} 
\caption{Observed (left panels) and PM decontaminated (right panels) CMDs for the globular cluster Patchick 99 , employing the near-IR VVV (top panels) and optical Gaia DR 2 data (bottom panels).}
\label{cmdpat}
\end{figure}

Successively,  we calculate the mean cluster PMs (adopting the $\sigma$-clipping method) measured from Gaia~DR~2, yielding $\mu_{\alpha}=(-2.98 \pm 1.74)$ mas~$yr^{-1}$ and $\mu_{\delta}=(-5.49 \pm 2.02)$ mas~$yr^{-1}$, which are in good agreement with the PM values found by the GMM method. After that, we selected as cluster members only stars within $3\ mas\ yr^{-1}$, following the procedure of \cite{Garro_2020}, indicated by the black circle drawn in Figure \ref{pm_Pat99}. Also, the size of the circle is set taking into consideration the typical size of GC members in the VPM diagram for other well studied MW bulge GCs. Additionally,  in Figure \ref{pm_cont} we show the PM distributions in a wide range ($-20<\mu_{\alpha}<15$ mas yr$^{-1}$ and $-20<\mu_{\delta}<10$ mas yr$^{-1}$),  considering only stars situated within $3'$ from the Patchick 99  centre (blue histogram), attempting to minimise field stars.  The resulting star sample after the PM cut is shown with the red histogram.  Each histogram is constructed considering the sum of all the combined values per unit bin, selecting a fixed bin size of $0.8$ mas yr$^{-1}$ (value that allowed us to visually distinguish the two peaks of the Gaussian distributions).  We can appreciate double peaked distributions for both blue histograms, with a very little separation (as expected from the GMM method),  suggesting that we have at least two overlapping populations: the bulge field stars and the star cluster.  Indeed, fitting two Gaussian distributions we assign a narrower and more peaked distribution for Patchick 99  (green curve) and a wider distribution for Bulge field stars (grey curve).  The total distribution (black curve) is obtained by coadding the two Gaussians, in order to fit the entire sample.  \\
Figure \ref {cmdpat} displays the PM decontaminated CMDs for VVV and Gaia data.  We can recognise that the main GC features, like RGB and RC, remain but, in comparison with the statistical decontamination (Figure \ref{statdec}), the foreground disk population is now better removed. \\

Following these procedures, we have obtained the decontaminated near-IR ($K_s$ versus $J-K_s$) and optical ($G$ versus $BP-RP$) CMDs for our cluster, as shown in the right panels of Figures \ref{cmdpat} and in Figure \ref{cmdiso_pat99}. We use those CMDs to derive the cluster’s main parameters.  Both diagrams clearly show the RC at $K_{s}=(12.5\pm 0.04)$ mag and $G=(15.5\pm 0.05)$ mag, suggesting that P99 is a metal-rich GC. It is crucial to use RC stars in this kind of studies, because these stars are very good distance indicators.\\

\subsection{Estimation of Physical Parameters for Patchick 99}
Nowadays, a wide range of reddening maps already exist, such as \cite{2011ApJ...737..103S}, \cite{2011A&A...534A...3G,2018MNRAS.481L.130G}, \cite{2018A&A...616A..26M}, \cite{10.1093/mnras/stz1752}, \cite{10.1093/mnrasl/slaa028}. We estimate the reddening and the extinction toward this GC following \cite{2018A&A...609A.116R} and we benefit from the clear position of the RC in the CMDs. We adopt the intrinsic RC magnitude $M_{K_s}=(-1.605\pm 0.009)$ mag and the intrinsic colour $(J-K_{s})_{0}=(0.66\pm 0.02)$ mag. Using the latter, we calculate a color excess of $E(J-K_s)=(0.12\pm 0.02)$ mag,  in accordance with the adopted reddening map, and the extinction $A_{K_s}=0.72\times E(J-K_s)= (0.09\pm 0.01)$ mag \citep{1989ApJ...345..245C}. Consequently, we use these values in order to determine distances, so we measure a distance modulus $(m-M)_{0}=(14.02\pm0.01)$ mag equivalent to a heliocentric distance $D=(6.4\pm 0.2)$ kpc. \\
Similarly, for optical wavebands we identify again RC stars and adopt the intrinsic RC magnitude $M_{G}=(0.459\pm0.009)$ mag \citep{2018A&A...609A.116R}. The near-IR extinction $A_{K_s}=(0.09\pm 0.01)$ mag corresponds to the optical extinction $A_G=0.86\times A_{V}=(0.68\pm 0.08)$ mag, deriving also $E(BP-RP)=A_{G}/3.12=(0.21\pm0.03)$ mag \citep{1989ApJ...345..245C,1994ApJ...422..158O}. Therefore, the resulting cluster distance modulus from the Gaia optical photometry is $(m-M)_{0}=(14.23\pm 0.1)$ mag and the distance $D=(7.0\pm 0.2)$ kpc,  agrees within 3$\sigma$ with our near-IR distance. \\

\begin{figure}[h]
\centering
\includegraphics[width=4.4cm, height=6.2cm]{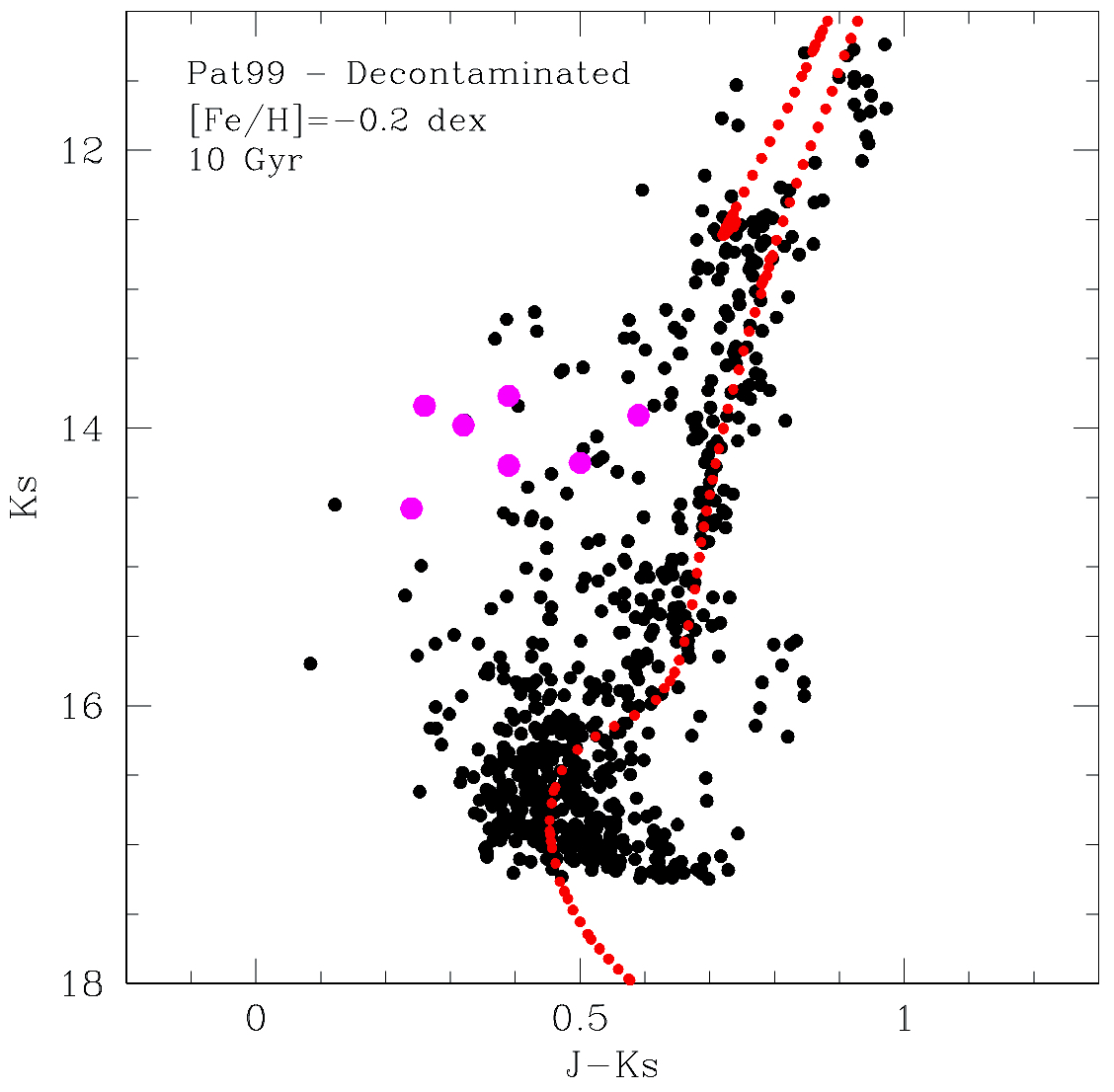} \includegraphics[width=4.4cm, height=6.2cm]{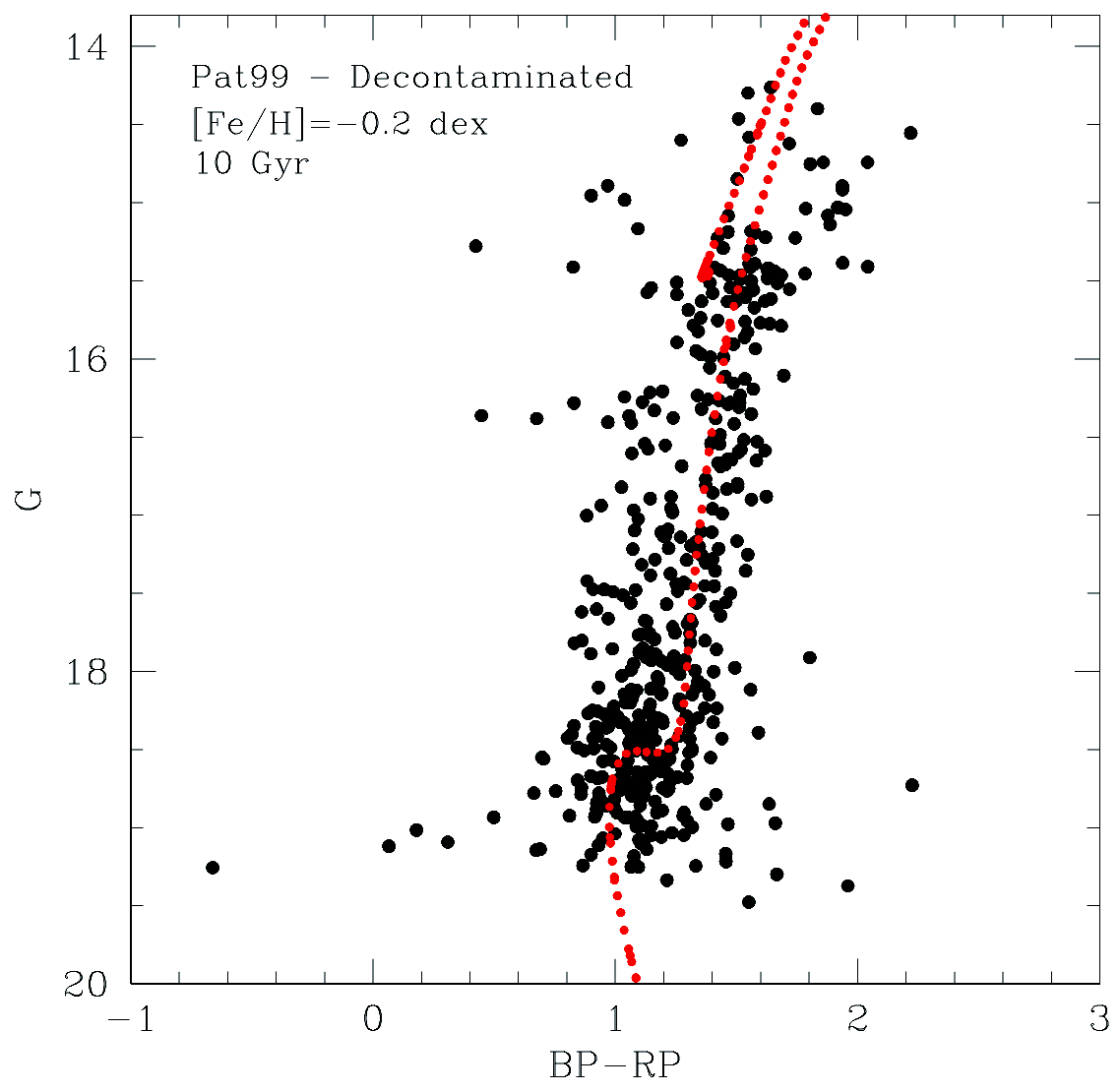}
\caption{The VVV near-IR (left panel) and Gaia DR 2 optical (right panel) PM decontaminated CMD for the globular cluster Patchick 99. The red dotted line is fitted by a PARSEC isochrone with age 10 Gyr and metallicity $[Fe/H] = -0.2$ dex. The magenta points represent the position of RR Lyrae star members in the near-IR CMD.}
\label{cmdiso_pat99}
\end{figure}

\begin{figure}[h]
\centering
\includegraphics[width=4.4cm, height=6cm]{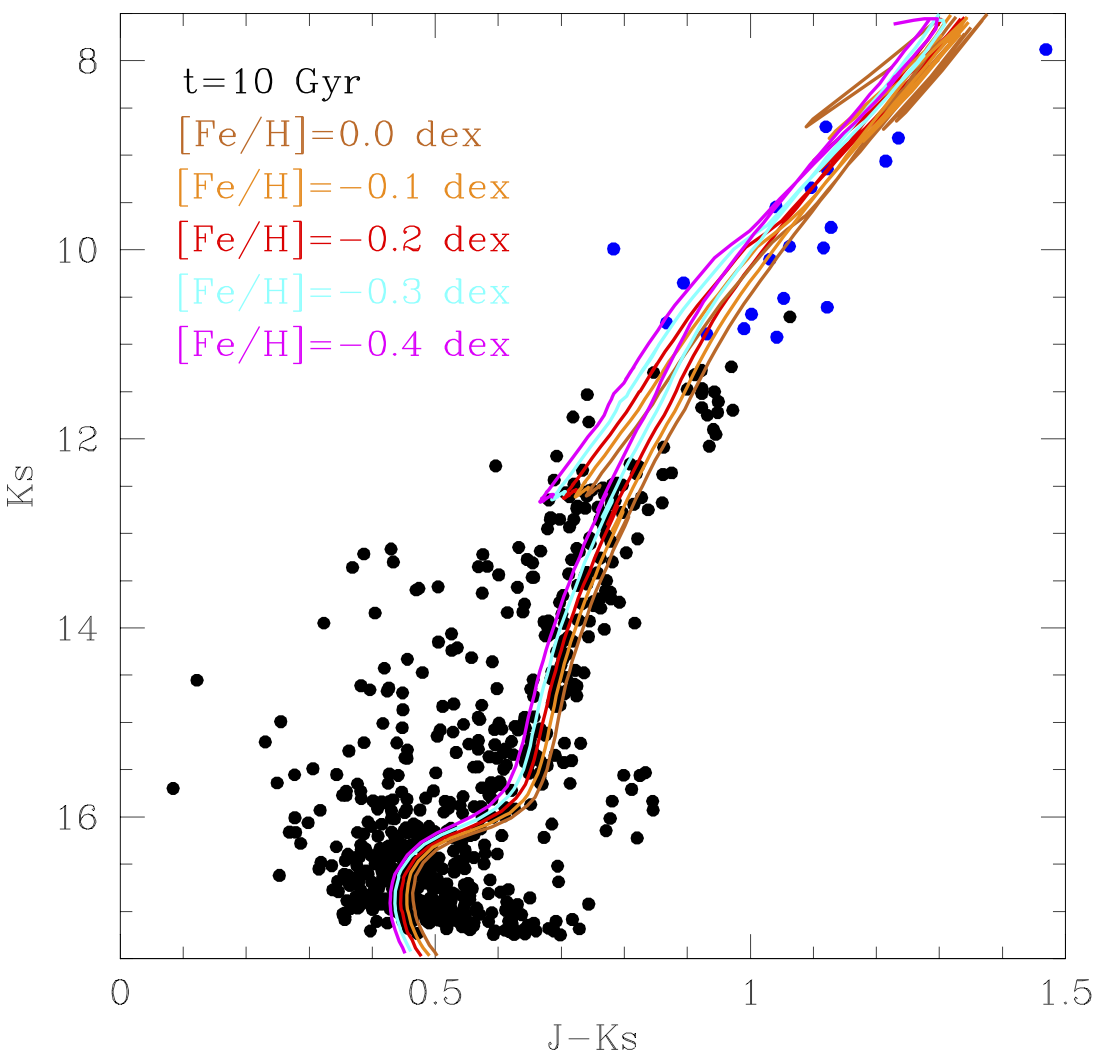} 
\includegraphics[width=4.4cm, height=6cm]{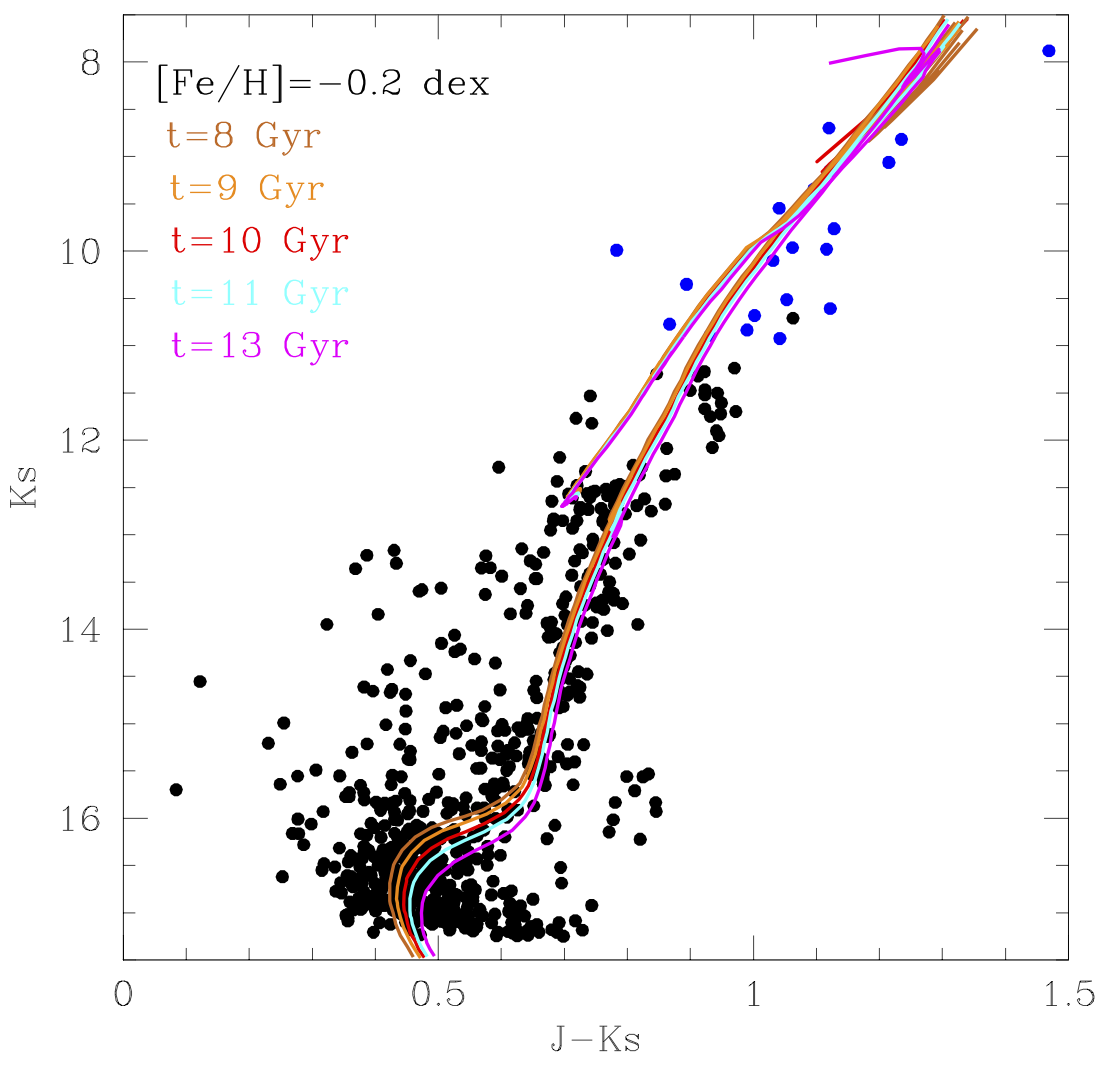}
\caption{Fitting of isochrones, in the VVV CMDs, at fixed metallicity of $[Fe/H]=-0.2$ dex and variation in ages t=8, 9, 10, 11, 13 Gyr (right panel); while at fixed age of t=10 Gyr and different metallicities $[Fe/H]=0,\ -0.1,\ -0.2,\ -0.3,\ -0.4$ dex (left panel). Black points represent the VVV cluster members, whereas blue points are those from 2MASS catalogue.}
\label{agemetal}
\end{figure}

Using our estimates of the reddening and distance we were able to fit PARSEC\footnote{http://stev.oapd.inaf.it/cgi-bin/cmd} isochrones \citep{2012MNRAS.427..127B, 2017ApJ...835...77M} to the CMDs of Patchick~99. This allows us to have estimation of the metallicity and the age of the GC, finding $[Fe/H] = (-0.2\pm 0.2)$ dex and $t=(10\pm 2.0)$ Gyr, respectively.  In order to obtain a more robust result, we have also derived the metallicity from the slope of the RGB ($\alpha=0.11\pm0.01$) using the calibration of \cite{10.1093/mnras/stw2435}, finding $[Fe/H]=-0.3 \pm 0.2$ dex, in agreement with our original estimate.  Figure \ref{cmdiso_pat99} shows a good fit of the isochrones with the above mentioned values of metallicity and age in both our optical and near-infrared CMDs. We determined the errors in age and in metallicity by changing the isochrones at a fixed metallicity and a fixed age, respectively,  until they do not fit the cluster sequence simultaneously in the near-IR and optical CMDs. Therefore,  it is clear that although age is one of the fundamental parameters to characterise GCs,  its correct determination is still one of the most complicated tasks to perform, especially in environments with high stellar density and with differential reddening, such as the Galactic bulge. This is mainly due to difficulties to reach the main sequence turn-off (MS-TO) point. We can appreciate that problem in Figure \ref{agemetal}, which displays the VVV and 2MASS PM decontaminated CMDs.  We include 2MASS cluster members with $K_{s}<11$ mag, as they are more sensitive to the metallicity variations (e.g. , \citealt{McQuinn_2019}). We can see that variations in metallicity have an especially strong effect along the RGB and RC,  and vice-versa, we can notice evident changes in age in the MS-TO region. For this reason our age should be taken as a crude estimate, since we are unable to measure it precisely. \\

All of the previously derived parameters,  including position, metallicity and kinematics confirm that Patchick 99  is a bulge GCs, but if we consider its age,  it could be even younger than typical metal-rich GCs present in the Galactic bulge, which have ages between 12 and 13 Gyr (e.g.,  \citealt{Carretta_2000}. However, many studies have demonstrated that GCs show a relatively wide range of ages in the bulge \citep{2003A&A...399..931Z,2011ApJ...735...37C, 2013A&A...559A..98V, 2018MNRAS.477.3507B}. \\

\begin{figure}[h]
\centering
\includegraphics[width=4.3cm, height=4.3cm]{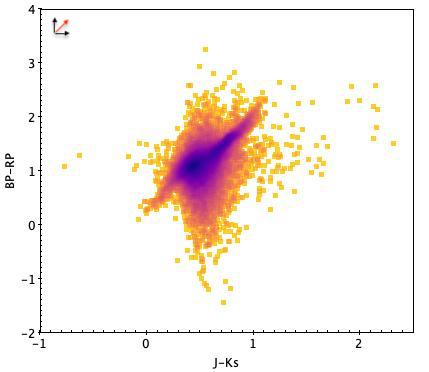}
\includegraphics[width=4.3cm, height=4.3cm]{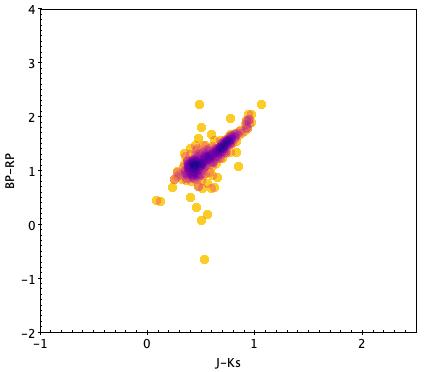}
\caption{Gaia+VVV optical-near-IR CCDs for Patchick 99, as density plot. On the left: \textit{squares} are the observed stars, using the entire catalogue without selection cuts; on the right: \textit{circles} represent star members of Patchick 99, considering bright stars previously selected. At the top-left of the panel, the red arrow represents the reddening vector.}
\label{ccdpat99}
\end{figure}

Figure \ref{ccdpat99} shows the Gaia+VVV optical-near-IR colour-colour diagram (CCD) for the whole catalogue without selection cuts, and for only cluster members of Patchick 99  with $K_s<15$ mag. We can clearly notice that the colour distribution reveals a tighter sequence for star members (circles, right panel) than for the field stars (squares, left panel).  In particular,  we can appreciate that there is a little differential extinction, in the right panel of Figure \ref{ccdpat99}. Based on the high resolution reddening map by \citealt{2020arXiv201002723S},  the mean reddening is $E(J-K_{s})=0.05$ mag in this field, which differs from our reddening value by $\Delta(E(J-K_s))=0.07$ mag at most. Hence, we believe that our estimation is in good agreement with the adopted map,  discarding this as a significant problem. \\

Finally, we derive the total luminosity of our cluster, using the catalogue with only cluster members. Once the flux for each star is measured, we derive the absolute magnitude from the total flux, obtaining $M_{K_s}=(-7.0 \pm 0.6)$ mag, equivalent to $M_{V}=(-4.5\pm0.8)$ (for a typical GC colour of $(V-K_{s})=(2.5\pm 0.5)$ mag), indicating that it is a very faint GC.  This value is an underestimate of the total luminosity, since the faintest stars are missing. Regarding this, we compared Patchick 99  luminosity with that other known GCs similar to Patchick 99  in metallicity: NGC\,6553 ($[Fe/H]=-0.18$ dex; $D=6.01$ kpc; $M_V=-7.77$ mag), NGC\,6528 ($[Fe/H]=-0.11$ dex; $D=7.9$ kpc; $M_V=-6.57$ mag) and Terzan\,5 ($[Fe/H]=-0.23$ dex; $D=5.9$ kpc; $M_V=-7.42$ mag).  Using the same method, and also spanning on the same magnitude range ($10.7 < K_s < 17.2$), we derive the integrated near-infrared absolute magnitude for each GC, finding $M_{Ks}\ (NGC\,6553)=-9.5$ mag, $M_{Ks}\ (NGC\,6528)=-8.0$ mag and $M_{Ks}\ (Ter5)=-9.6$ mag. We converted them in $V-$band and scaled to the $M_V$ values from the 2010 version of the  \cite{1996AJ....112.1487H} catalogue. Therefore, estimating the $M_V$ differences and assuming the similarities between the GCs, we derive that the total luminosity for Patchick 99  is $M_V=-5.2$ mag, $\sim 2.2$ less luminous of the MW GC luminosity function peak ($M_V=(-7.4\pm0.2)$ mag from \citealt{Harris_1991, ashman_1998}) \\

All these parameters are summarised in Table \ref{table1}.

\subsection{The derivation of the surface density profile}
A detailed description of the structural parameters of Patchick~99 would need a careful treatment of completeness as well as relatively large numbers of member stars from which a radial profile can be derived. Traditionally, King profiles \citep{1962AJ.....67..471K} have been fitted to a significant fraction of MW old clusters, with however more and more evidence that other profiles might provide better fits, especially when extragalactic or massive GCs are included (e.g., \cite{Mackey}, \cite{2019MNRAS.485.4906D}). Although rich clusters allow for model-dependent sizes and structural properties, our sample of member stars is not enough to adequately address a comparison among the different models available. Instead, a King function is assumed to be a good approximation for the radial distribution of the stellar density.\\

\begin{figure}[h]
\centering
\includegraphics[width=7cm, height=5cm]{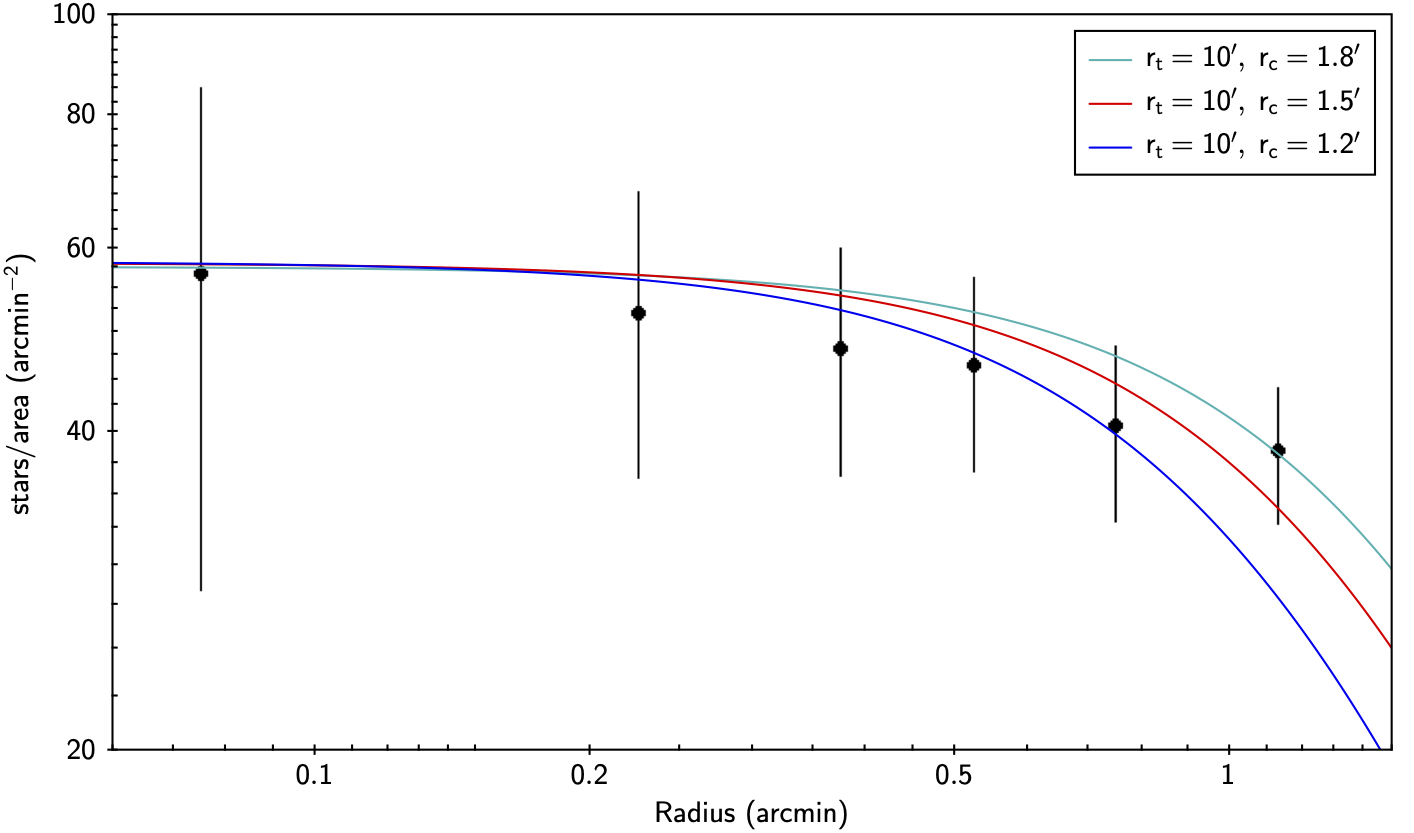}
\caption{Radial surface density of Patchick 99  (black points), taking into consideration only PM selected member stars. The coloured lines are King model profiles that best reproduce observational data, with a fixed value of the tidal radius $10'$ and different values of core radius, in cyan $r_c=1.8'$, in red $r_c=1.5'$ and in blue $r_c=1.2'$ Poisson errors are shown as error bars.}
\label{sizepat}
\end{figure}

The correct determination of the cluster centre has a fundamental importance, since an error in its position can lead to altered results regarding the density profile. For this reason, we have used the K-Means algorithm in Python to determine the centre in a robust way from our catalogue of cluster member stars. In summary, this is a machine-learning algorithm that searches for subgroups (or literally "clusters") in a given parameter space among a sample of data points. We checked visually that the resulting centres were close to the expected density peaks and we found that the \textit{initial} position ($R.A.=18h15m47s$ and $DEC=-29d48m46s$ (J2000)) coincides almost-perfectly with the resulting values of the centre, with a small separation of $\Delta RA\approx 2''$ and $\Delta DEC\approx 5''$.\\
Starting from these new coordinates, we divided our sample of member stars into ten circular annuli, out to a radius of $1.5'$, in steps of $0.15'$.  We do not need to apply geometric corrections as the resulting circular annuli are within the FOV of our catalogue. We performed the radial density profile of Patchick 99  (see Figure \ref{sizepat}) dividing the total number of stars in each radial bin by the area of that ring in order to derive the density in each bin. Poisson errors are also shown. We note that the error bars are relatively large, mainly due to the low-number statistics in each radial bin, especially in the innermost point (see Figure \ref{sizepat}). No background has been subtracted since we are dealing with a list that is expected to be free from foreground stars and other contaminants.\\

We have overplotted a family of King profiles adopting different values of core radius ($r_{c}=1.8'$, $1.5'$, $1.2'$), but keeping the tidal radius fixed at $10'$. The choice to use that value of tidal radius is justified following \cite{Boer_2019} (see their Figures 13 and 14), where a clear correlation between the tidal radius and galactocentric radius of a given GC is illustrated. As shown in Figure \ref{sizepat},  the observational points are best represented by $r_{c}=1.2'$,  which is equivalent to $r_{c}\sim 1.1$ pc.  Also, we consider the tidal radius $r_{t}=10'$ as the size of the cluster, which corresponds to a physical size of $9.3$ pc.

\subsection{The RR Lyrae stars in Patchick 99}
RR Lyrae variable stars represent a very old-population in our Galaxy, found in the Galactic halo and bulge. In particular,  according to \cite{Pietrukowicz_2015} their spatial density ranges between 80 and 440 RRL/deg$^2$ in the bulge. \\
They are also considered powerful tools to reveal the nature of a GC candidate because if detected in a stellar cluster, RR Lyrae stars guarantee that it is an old GC.  Although these variables are used as excellent tracers of old and metal-poor populations, there are observational evidences that suggest the presence of RR Lyrae also in metal-rich GCs, such as NGC~6388 ($[Fe/H]=-0.44$ dex) and NGC~6441 ($[Fe/H]=-0.46$ dex) \citep{2002AJ....124..949P, 2005ApJ...630L.145C}, and NGC~6440 $[Fe/H]=-0.36$ dex), all of which are somewhat more metal-rich than Patchick 99.
Additionally, they are good reddening and distance indicators. \\

For these reasons, we have searched for these variable stars and we have detected 21 RR Lyrae stars within 12 arcmin from the Patchick 99  centre, matching the OGLE \citep{2014AcA....64..177S}, the VVV \citep{cite-key} and the Gaia \citep{2019A&A...622A..60C} catalogues. They are listed in Table \ref{rrlpat}, where we summarised: identification, location, period, photometry in different filters, heliocentric distance, distance from the centre of the GC,  type for each source and their PMs. \\

\begin{figure}[h]
\centering
\includegraphics[width=7cm, height=7cm]{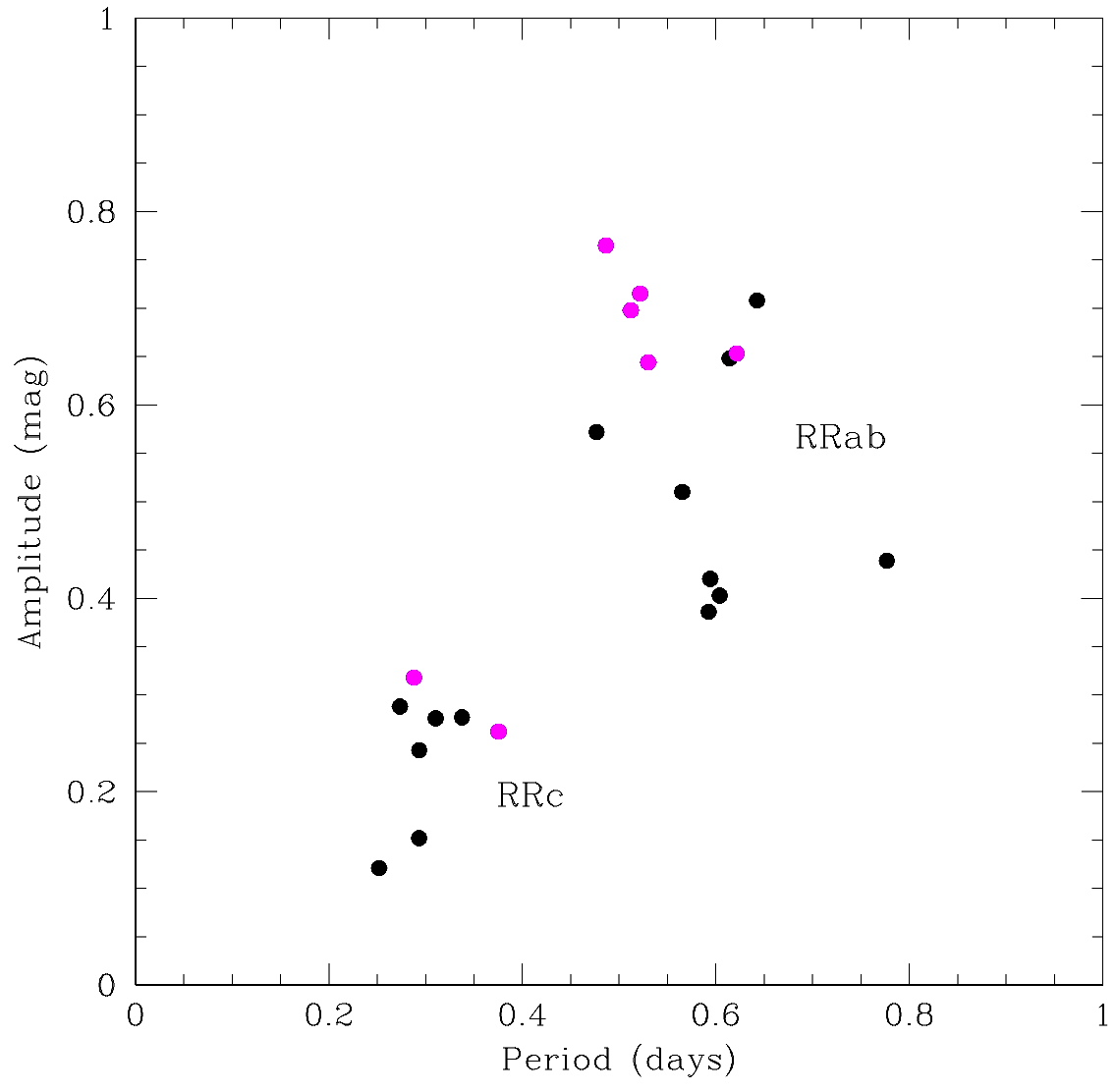}
\caption{Bailey diagram for detected 21 RR Lyrae stars (black points), situated within 12 arcmin from the Patchick 99  centre, while we highlight the RR Lyrae member stars with magenta points.}
\label{bd}
\end{figure}

Initially we constructed their phased light curves, in order to discriminate their pulsation mode. We identify 13 fundamental mode pulsators (ab-type RR Lyrae stars) with long periods ($P\gtrsim0.4$ days) and large amplitudes, and 8 first-overtone pulsators (c-type RR Lyrae stars), with short period ($P\lesssim0.4$ days) and small amplitudes. The RRc and RRab pulsators also differ in the morphology of their light curves, since the RRab's have typically asymmetrical light curves, unlike RRc's.  Figure \ref{bd} displays the Amplitude-Period diagram (so-called Bailey diagram, \citealt{1919AnHar..78..195B}) in such a way to highlight that differences.  Indeed, we consider the Bailey diagram a solid tool for verifying our estimations.

\begin{figure}
\centering
\includegraphics[width=7cm, height=7cm]{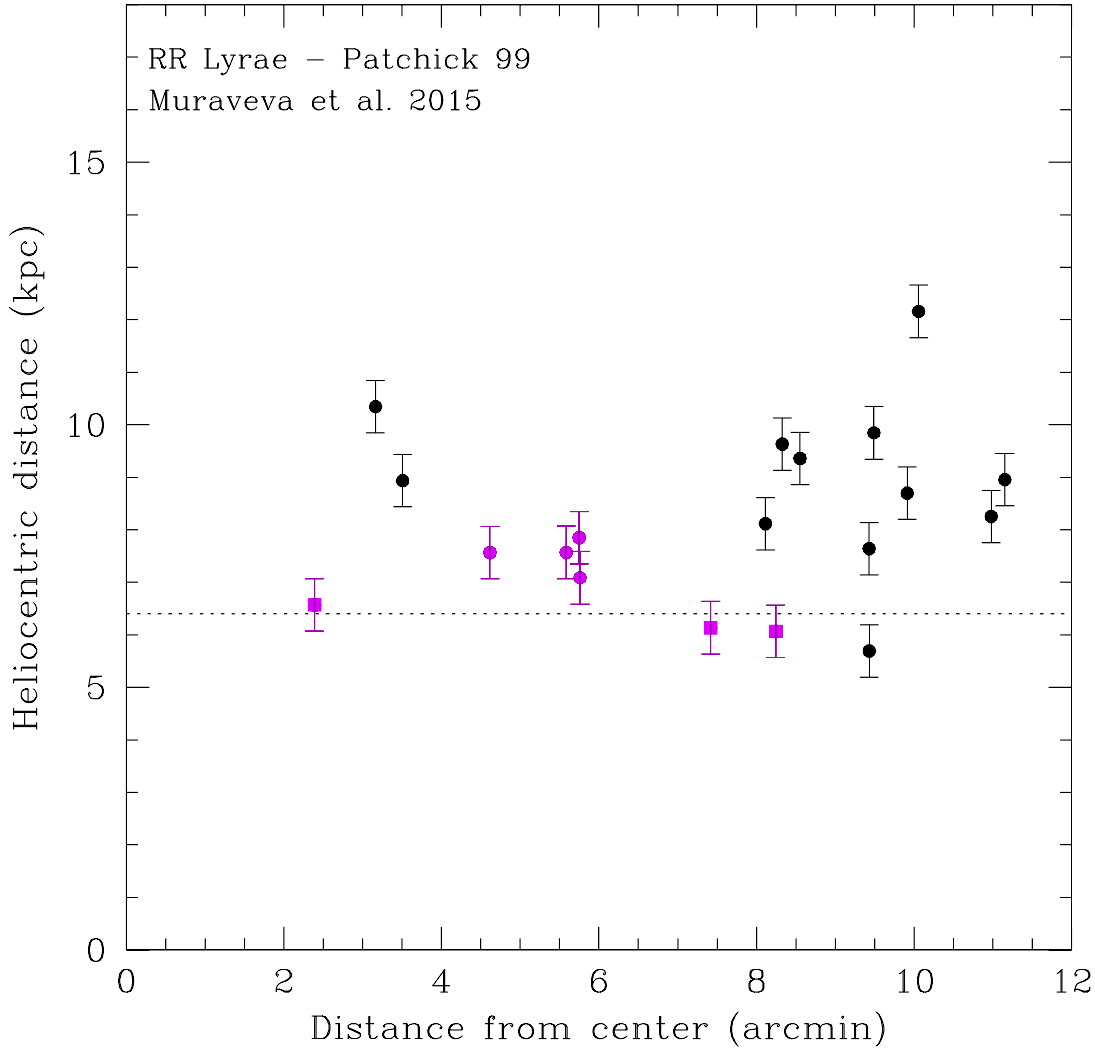}
\caption{Spatial distribution of RR Lyrae variable stars detected within $12'$ from the centre of Patchick 99. The magenta points represent the seven RR Lyrae stars considered cluster members,  whereas we highlight the PM members with magenta squares. The distance errorbars are of 0.5 kpc for each point. The dotted line depicts the heliocentric distance at $D=(6.4\pm0.2)$ kpc derived from the VVV near-IR photometry.}
\label{distanceRRL}
\end{figure}

\begin{figure}[h]
\centering
\includegraphics[width=7cm, height=7cm]{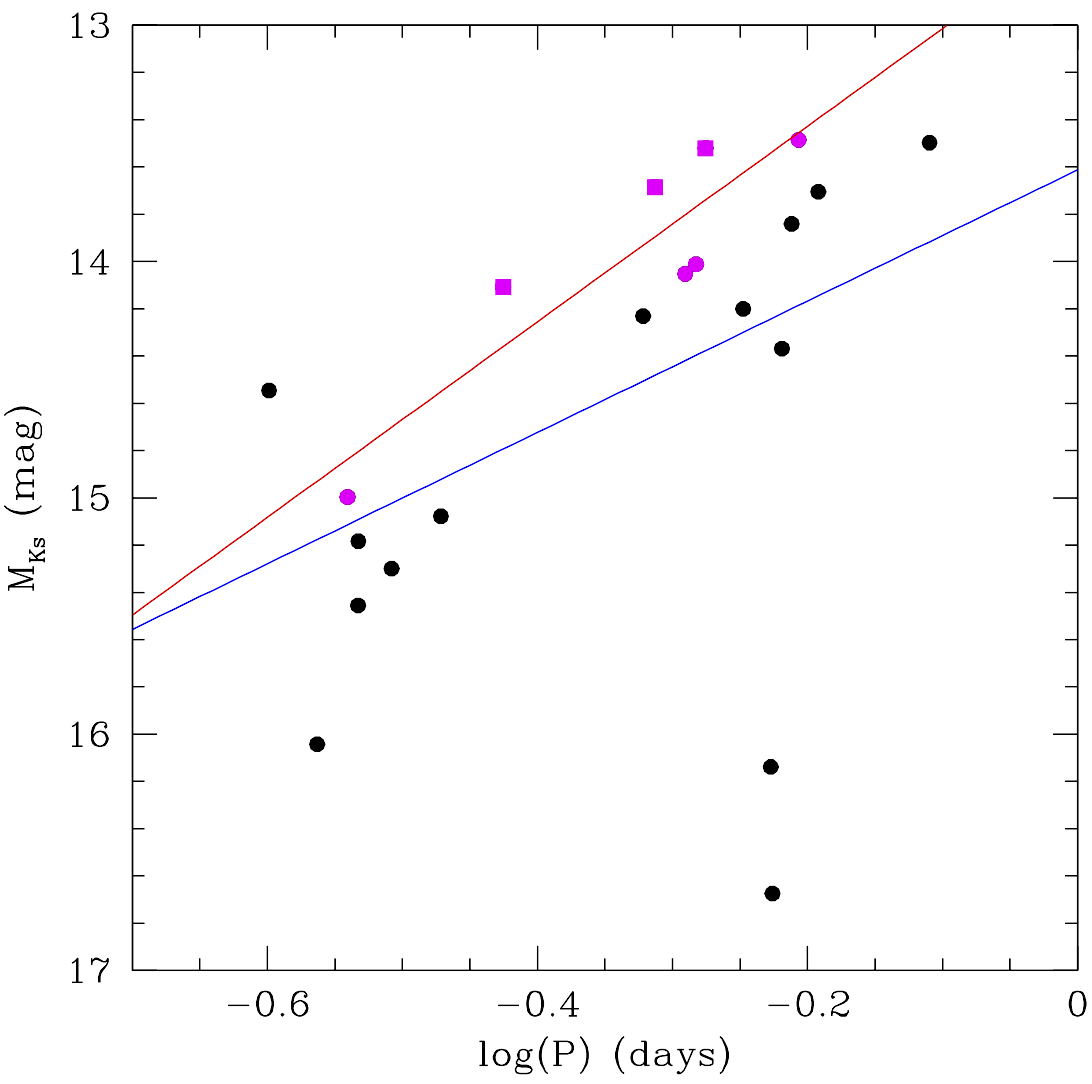}
\caption{$M_{Ks}-\log(P)$ diagram for all RR Lyrae sample (black points) and RR Lyrae star members (magenta points, while the squares represent the PM variable members). Blue and red lines represent the least-squares fit for all RR Lyrae sample and cluster members, respectively.}
\label{MksP_diagr}
\end{figure}

\begin{figure}
\centering
\onecolumn
\includegraphics[width=8cm, height=5.5cm]{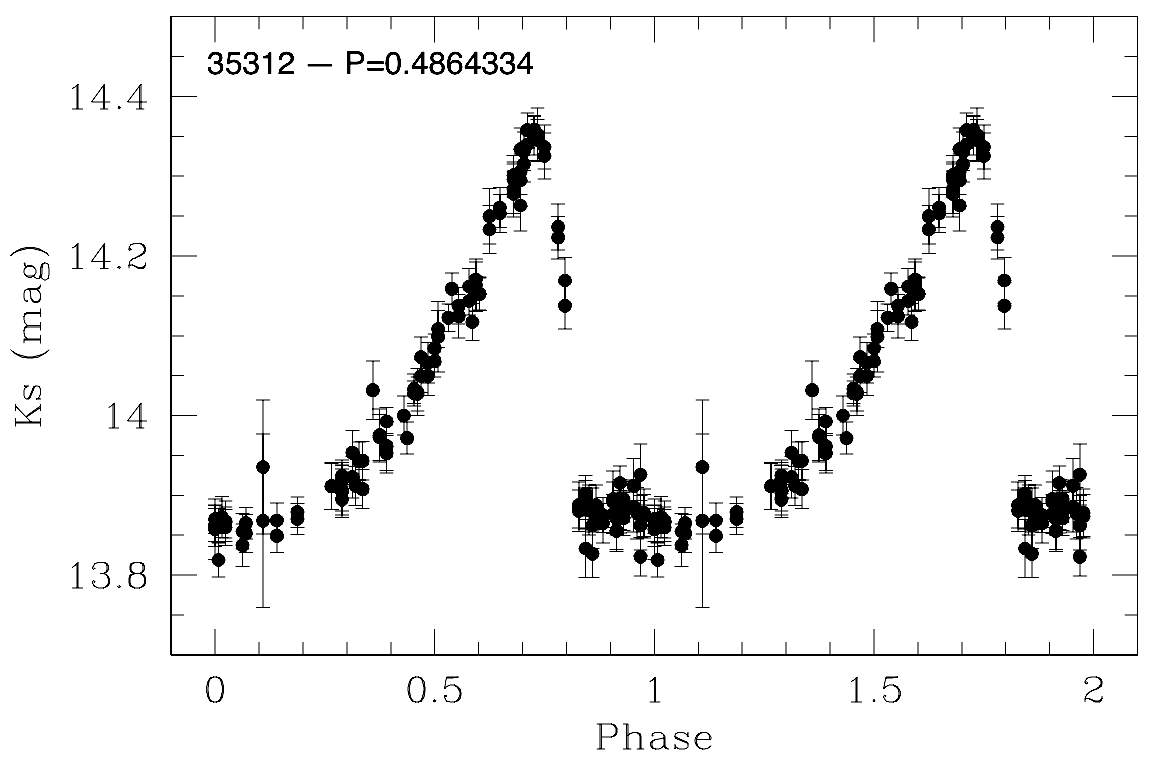}
\includegraphics[width=8cm, height=5.5cm]{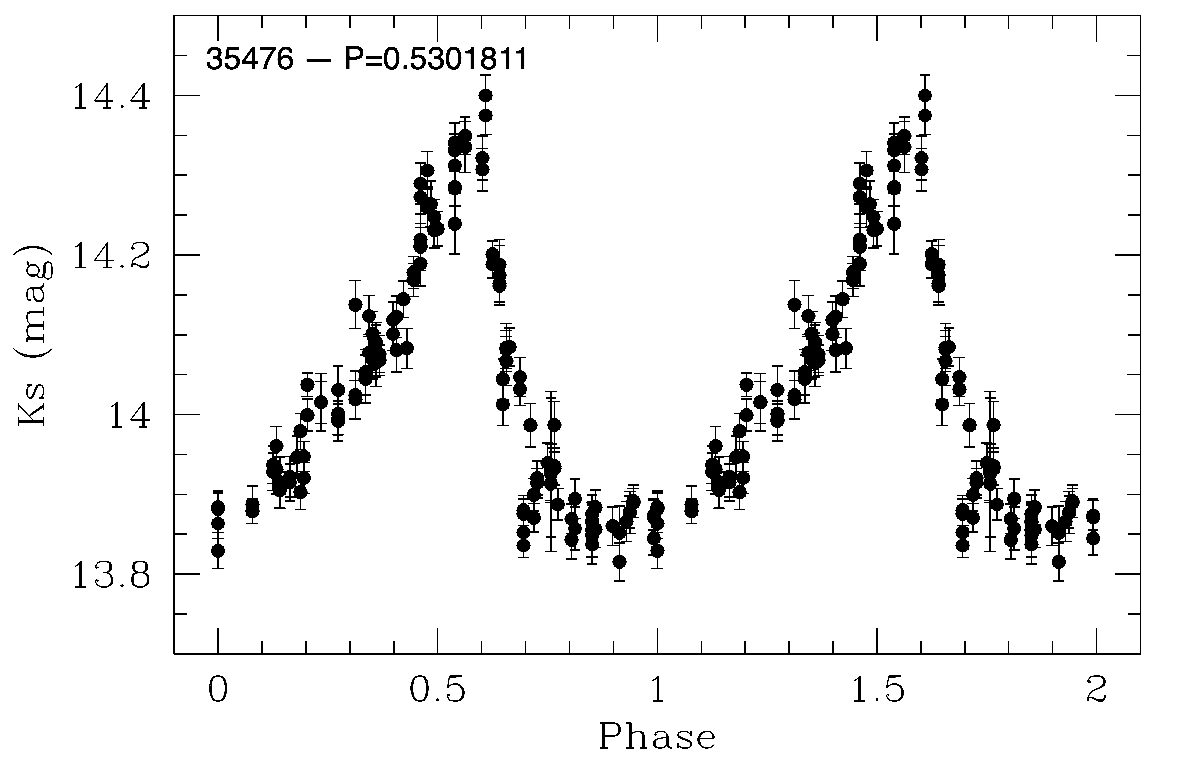}
\includegraphics[width=8cm, height=5.5cm]{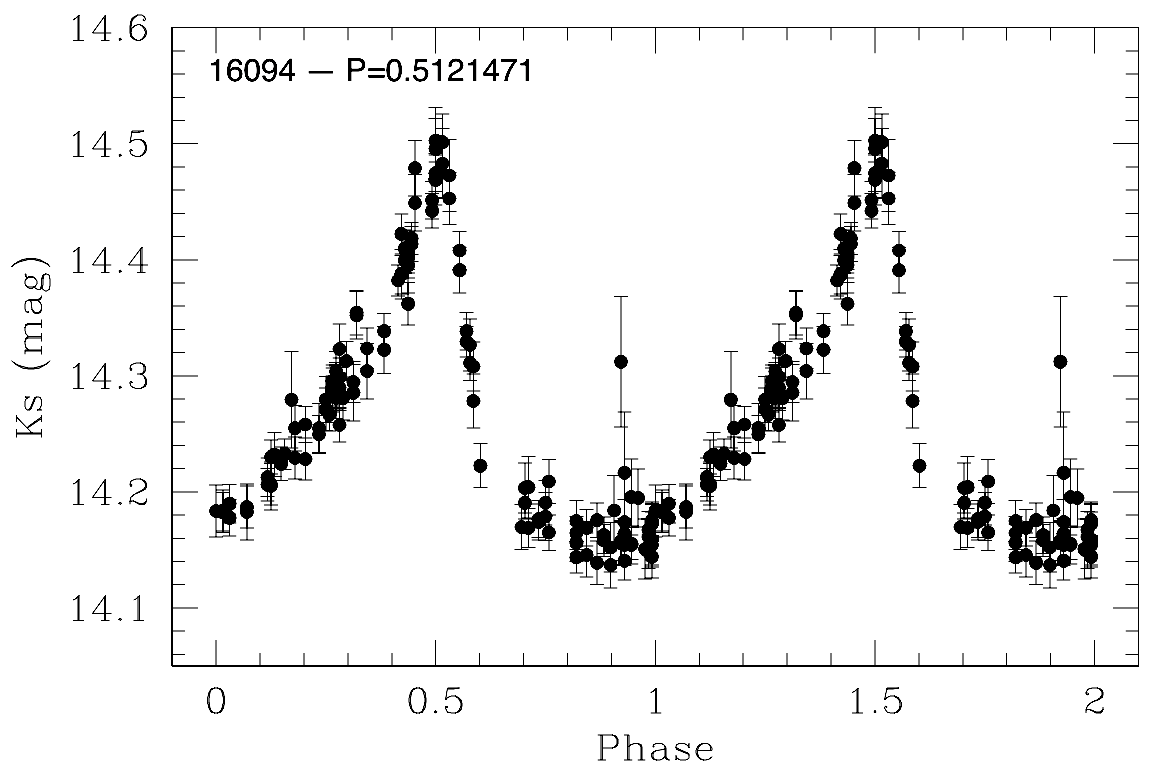}
\includegraphics[width=8cm, height=5.5cm]{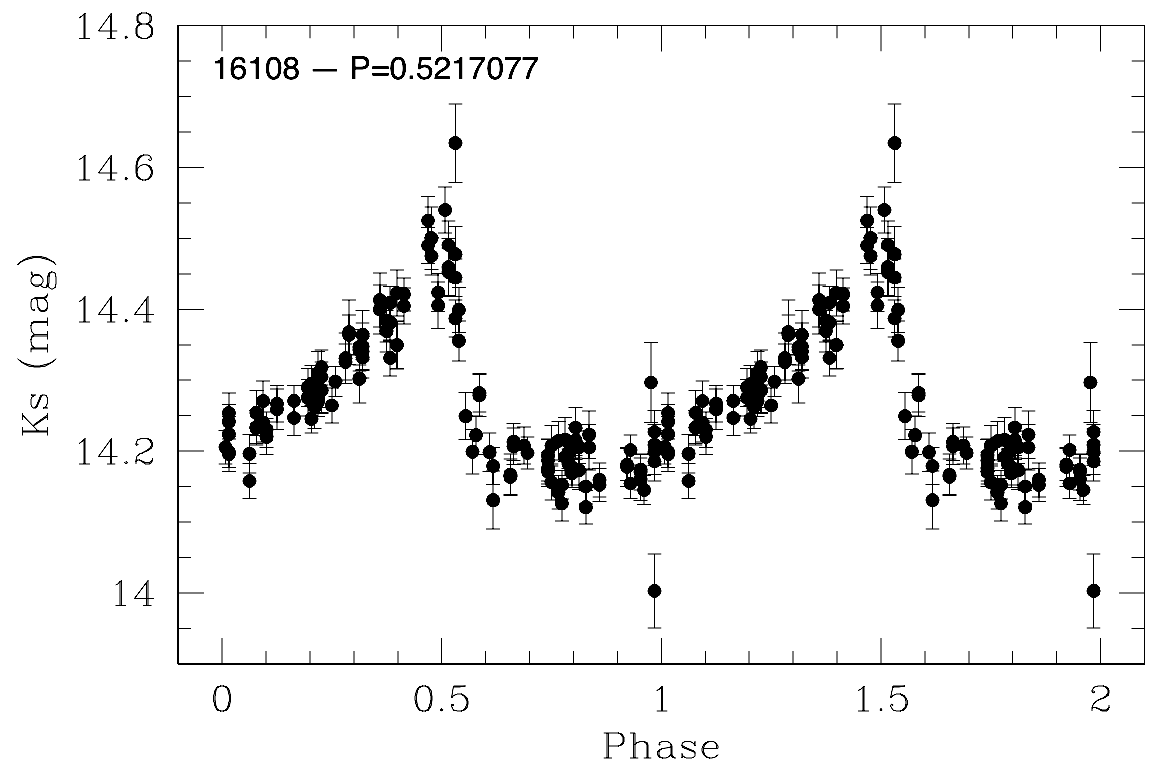}
\includegraphics[width=8cm, height=5.5cm]{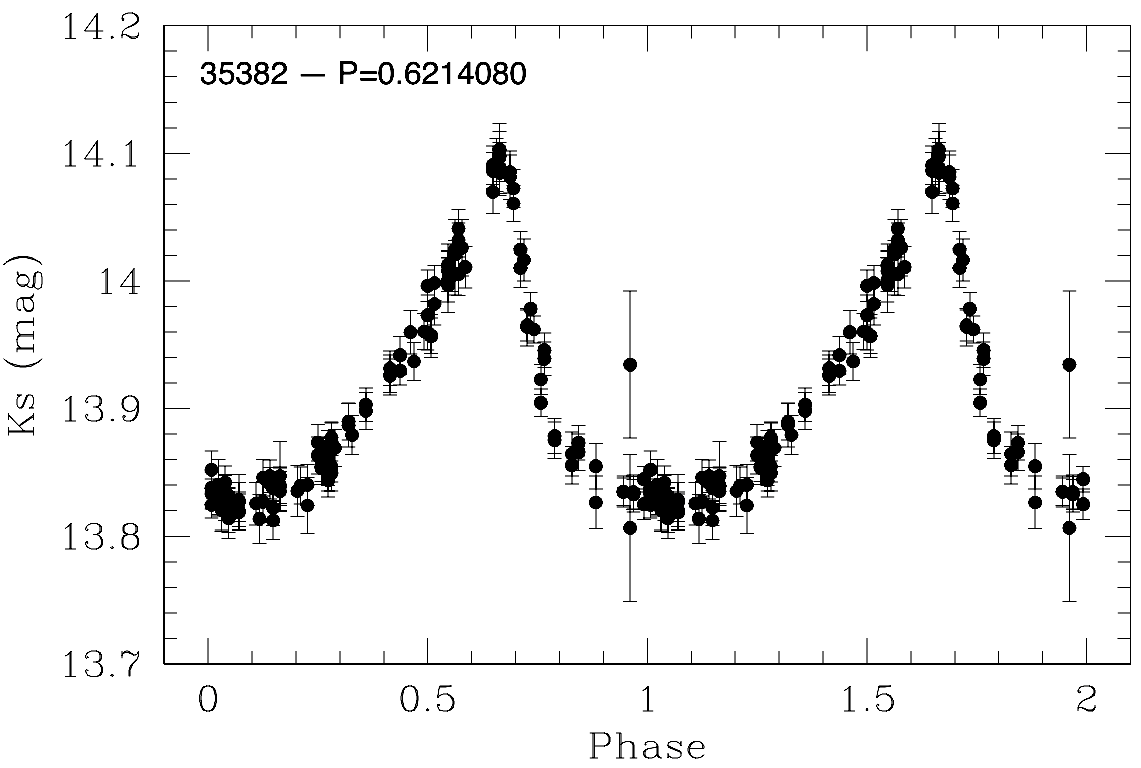}
\includegraphics[width=8cm, height=5.5cm]{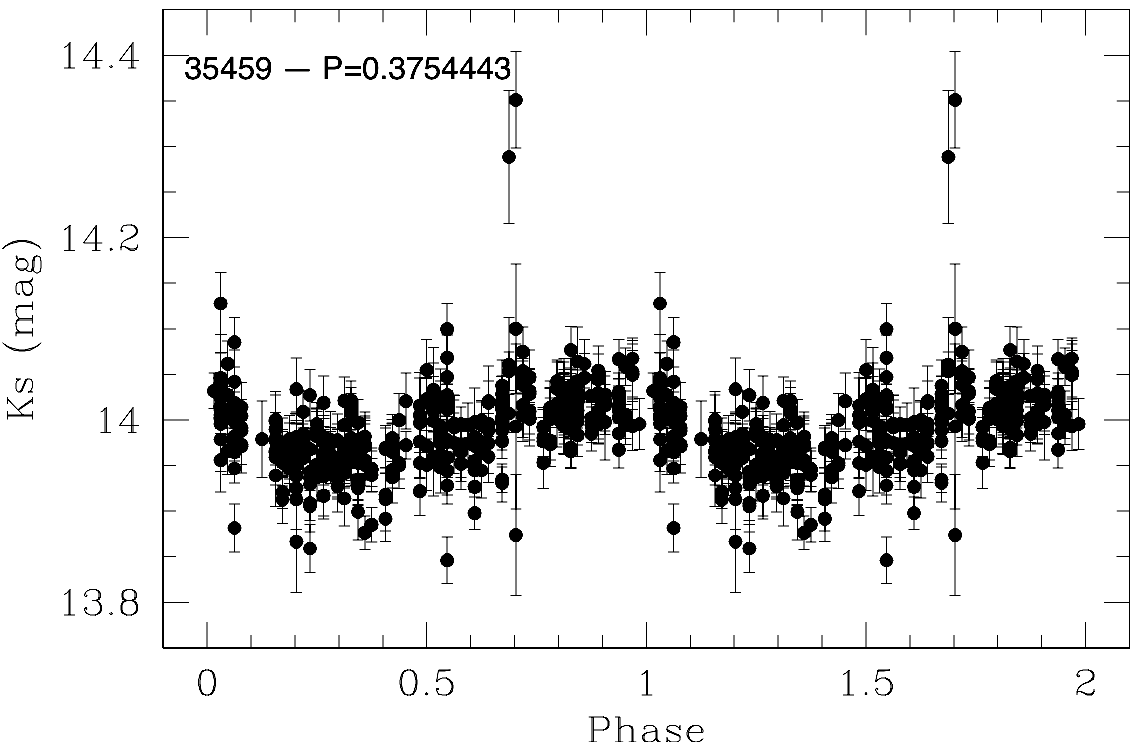}
\includegraphics[width=8cm, height=5.5cm]{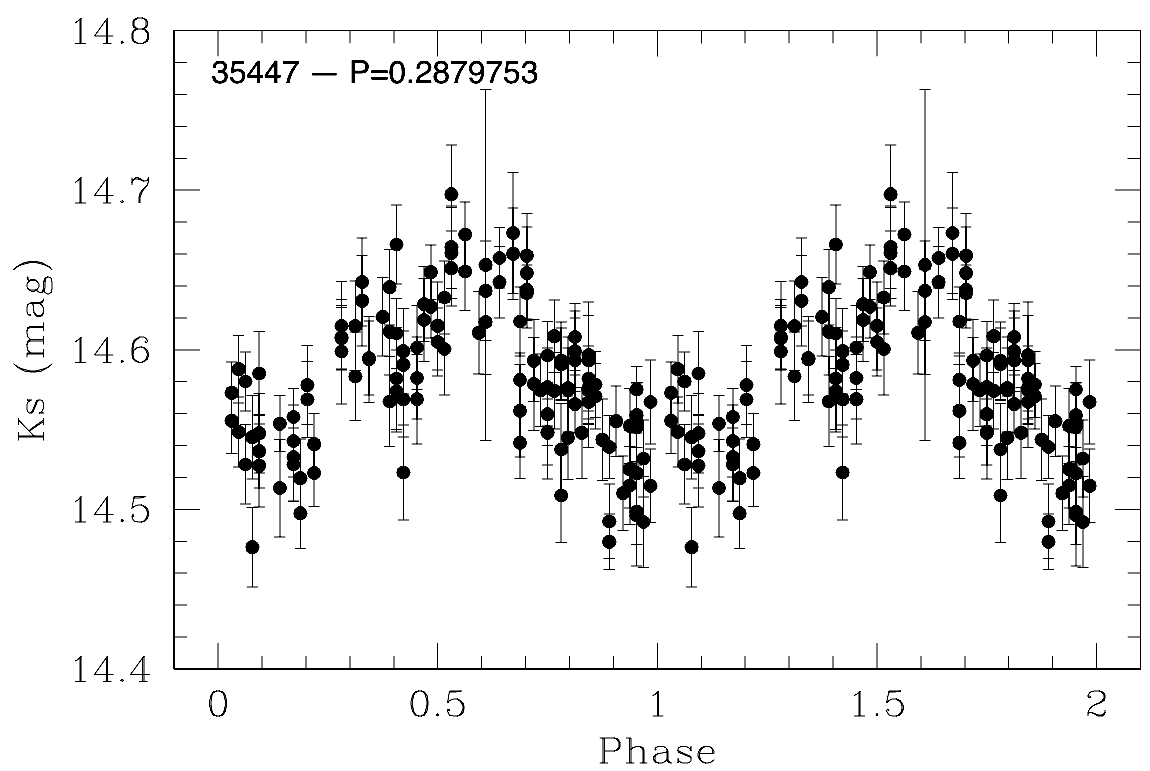}
\caption{Phased light curves for the RR Lyrae star members of Patchick 99 GC, classified as ab-type with $P>0.4$ days and c-type with $P<0.4$ days.}
\label{lcRRL}
\end{figure}

Our main goal is to identify which RR Lyrae stars are real members of Patchick 99.  Firstly,  we measured their projected distance from the centre using the centroid of each variable stars. Based on these distances and considering the size of the cluster, we believe that seven RR Lyrae stars are members of Patchick 99, as shown in Figure \ref{distanceRRL} and evidenced by a star symbol in Table \ref{rrlpat}. In addition, we compare their projected distance with the resulting heliocentric distance (Figure \ref{distanceRRL}),  confirming again the membership.  In detail, we used the period-luminosity (PL) relation in $K_s$-band following \cite{Muraveva_2015} (with RRc stars “fundamentalized” by adding $0.127$ to the logarithm of the period) in order to derive their heliocentric distance. Once selected the RR Lyrae members, we estimate the mean distance, finding a value of $D=7.06\pm 0.48$ kpc,  consistent with the heliocentric distances obtained from the VVV and Gaia photometries (see Section 4), since these values differ by less than 2$\sigma$. 
On the other hand, we discard the others as members since they have heliocentric distances larger than those found using VVV near-IR photometry,  or they are located outside our assumed tidal radius.  However, we can restrict the RR Lyrae sample to the ones that have PMs matching the mean GC PM within the errors, we find that only 3 of them fulfil this requirement: OGLE-BLG-RRLYR-35312, 35476 and 35459, favouring a short distance of $6.3\pm 0.5$ kpc in the mean.
Additionally, Figure \ref{distanceRRL} does not show OGLE-BLG-RRLYR-35364 and OGLE-BLG-RRLYR-35381, since their distance values are very large and probably they are located in the Galactic halo or in the Sagittarius dwarf galaxy. 
We also tested two other PL relations from \cite{2017A&A...605A..79G} and \cite{2017A&A...604A.120N}, yielding slightly larger distance values. However, we use the comparison between the three PL relations, in order to assign a mean distance error of 0.5 kpc. Once absolute magnitudes are obtained, we also build-up the PL~relation diagram (Figure \ref{MksP_diagr}). This allowed us to confirm again the membership, since we find a stronger relation when the least-square fit is used on the star cluster members (red line),  rather than on the whole sample (blue line).  We also show the VVV phased light curves for each RR Lyrae member (Figure \ref{lcRRL}). We note that the light curve profile for OGLE-BLG-RRLYR-35459 is slightly worse than the others,  since this variable appears a bit blended. \\
Moreover, we derive the expected spatial density from \cite{Pietrukowicz_2015} and \cite{2020arXiv201006603N}.  Specifically, following \cite{2020arXiv201006603N}, we are able to predict a mean density of 4 field RR Lyrae type-ab in the same field of Patchick 99. We found 21 variables in total, including 13 RRab.  Hence, this value represents a clear excess over the background ($ \sim 4.5 \sigma $ detection).

Finally, the magenta points in Figure \ref{cmdiso_pat99} define the position of the seven RR Lyrae stars in the VVV CMD. Their near-IR magnitudes are on average much fainter than the RC ($\sim 1 - 1.5$ mag).  However, both theory and observations suggest that the RR Lyrae indeed should be fainter than the RC in the near-IR CMDs and also indicate that more metal-rich RR Lyrae are fainter than metal-poor ones.  For example, \cite{2004A&A...426..641C} explored the predicted behavior of the pulsators as a function of the horizontal branch morphology and over the metallicity range Z=0.0001 to 0.006, finding that the RR Lyrae luminosity decreases with increasing metallicity.  Further, \cite{Marconi_2018} constructed new sets of helium-enhanced (Y = 0.30, Y~=~0.40) nonlinear, time-dependent convective hydrodynamical models of RR Lyrae stars covering a broad range in metal abundances (Z = 0.0001 -- 0.02), finding that an increase in helium content from the canonical value (Y = 0.245) to Y = 0.30 -- 0.40 causes a simultaneous increase in stellar luminosity and in pulsation period (see their Figure 1 and 2).  \\
Also, from the observational point of view,  we have compared the location of our RR Lyrae with those confirmed PM members of the cluster NGC~6441 from Alonso-García et al.  (in preparation). The RR Lyrae stars in Patchick 99 fall in the expected region of the near- IR CMD for this metal-rich cluster. In any case, spectroscopic metallicities are needed to confirm if this is the most metal-rich GC known with RR Lyrae.


\section{Summary and Conclusions}
We have investigated the GC candidates Patchick 99 and TBJ 3 in order to understand their real nature. \\
They are analysed in optical and near-IR wavelengths with the goal of determining their  physical parameters. The visual examination of TBJ 3 at various wavelengths confirms that it is not a GC but from its mid-IR emission we conclude that TBJ 3 is a background galaxy. \\
Therefore, we have focused our work on Patchick 99. We have measured its reddening, extinction, distance modulus and heliocentric distance, size, proper motion, total luminosity, metallicity and age. We summarised all those values in Table \ref{table1}.\\
Even though the field contamination is severe, we used two different methods (statistical and PM decontamination) recovering the same results, with the GC RGB and RC clearly present. We conclude that Patchick 99 is a metal-rich GC,  located in the Galactic bulge at a heliocentric distance of $D= 6.6\pm 0.6$ kpc,  obtained as an average distance of three-independent methods (Gaia DR 2 and VVV photometries, and three RR Lyrae star members). We define it as a genuine GC because both CCD and CMDs yield a population with an age of 10 Gyr. These results are confirmed by the presence of seven RR Lyrae stars, located at $d<12'$ from the cluster centre and at the same distance of the cluster from the Sun.  However,  analysing their PM we can confirm that three of them are surely member of our cluster: OGLE-BLG-RRLYR-35312, 35476 and 35459, since their PMs match the mean cluster PM within the errors. Patchick 99 may be the most metal-rich globular cluster with RR Lyrae, to be confirmed by spectroscopic observations.
As mentioned previously, even if position, metallicity and kinematics confirm that it is a bulge GC, the age of Patchick 99 is a few Gyr younger than the typical age of metal-rich GCs ($\sim 12-13$ Gyr) in the same regions. In any case, it is even important to underline that our estimation of age is quite rough and we need deeper observations to be more accurate.\\
Another important conclusion is that Patchick 99 is a low-luminosity GC ($M_{Ks} = -7.0$ mag), suggesting that many other faint GCs are still to be revealed in the bulge of the Milky Way.

\begin{acknowledgements}
We gratefully acknowledge the use of data from the ESO Public Survey program IDs 179.B-2002 and 198.B-2004 taken with the VISTA telescope and data products from the Cambridge Astronomical Survey Unit. ERG acknowledges support from an UNAB PhD scholarship. D.M. acknowledges support by the BASAL Center for Astrophysics and Associated Technologies (CATA) through grant AFB 170002. D.M. and M.G. are supported by Proyecto FONDECYT No. 1170121.
J.A.-G. acknowledges support from Fondecyt Regular 1201490 and from the ANID’s Millennium Science Initiative $ICN12_{\_} 009$, awarded to the Millennium Institute of Astrophysics (MAS). 
This work has made use of the University of Hertfordshire's high-performance computing facility.
\end{acknowledgements}

\bibliographystyle{aa.bst}
\bibliography{bibliopaper}

\newpage
\begin{table}[h]
\centering 
\onecolumn
\caption{Main physical parameters for Patchick 99 determined using near-IR and optical datasets.}
\begin{tabular}{lc}
\hline\hline
\textbf{Physical parameters} & \textbf{Patchick 99} \\
\hline
RA (J2000) & 18h15m47s  \\
DEC (J2000) & -29d48m46s  \\
Latitude & $ 2^{\circ}.4884 $\\
Longitude  & $ -6^{\circ}.1452 $\\
$\mu_{\alpha}$ [mas/yr]& $-2.98 \pm 1.74 $ \\
$\mu_{\delta}$ [mas/yr]& $-5.49 \pm 2.02$ \\
$A_{K_s}$ [mag]& $ 0.09\pm 0.01$ \\
$A_{G}$ [mag]&  $ 0.68\pm0.08$  \\
$E(J-K_{s})$ [mag]& $ 0.12\pm 0.02$  \\
$E(BP-RP)$ [mag]& $ 0.21\pm 0.03$  \\
$(m-M)_{0}$ [mag] & $ 14.02\pm0.04 $\\
$D_{mean}$ [kpc] &$ 6.6\pm 0.6 $\\
$M_{K_s}$ [mag]& $-7.0\pm0.6$ \\
$M_{V}$ [mag] & $-5.2$ \\
$[Fe/H]$ [dex]& $ -0.2\pm0.2$ \\
Age [Gyr]&  $10.0\pm 2$  \\
$r_c$ [arcmin] & 1.2 (1.1 pc)\\
$r_t$ [arcmin] & 10 (9.3 pc)\\
\hline\hline
\end{tabular}
\label{table1}
\end{table}

\begin{table}
\onecolumn
\centering 
\caption{GMM analysis for each sub-sample (field stars, position  and PM selection)  applied to PM in RA.  The subscripts $1$ and $2$ indicate the first and second mode of the Gaussian distribution.}
\begin{tabular}{l|cc|cc|c}
\hline\hline 
                  & $\mu_{1}$ & $\sigma_{1}$ & $\mu_{2}$ & $\sigma_{2}$ & D  \\
          & [mas yr$^{-1}$] &  [mas yr$^{-1}$] &  [mas yr$^{-1}$] &  [mas yr$^{-1}$]&   \\
\cline{2-6}
Field stars &$-2.11\pm 0.11$ & $4.41\pm 1.98$& $-1.26\pm 0.52$&$13.99 \pm 1.98$  &$0.08\pm 0.05 $ \\
Position sel. &$-1.96 \pm 0.58 $& $4.08 \pm 2.8$& $-2.54\pm 0.55$&$10.71 \pm 3.16$ &  $0.07\pm 0.06$ \\
PM selection & $-3.35\pm 0.13$ & $1.42\pm 0.05$ &$ -0.73 \pm 0.14$ & $0.53\pm 0.08$ &  $2.46\pm 0.09$ \\
\hline\hline
\end{tabular}
\label{GMM_ra}
\end{table}

\begin{table}
\onecolumn
\centering 
\caption{GMM analysis for each sub-sample (field stars, position  and PM selection)  applied to PM in DEC.  The subscripts $1$ and $2$ indicate the first and second mode of the Gaussian distribution.}
\begin{tabular}{l|cc|cc|c}
\hline\hline 
                  & $\mu_{1}$ & $\sigma_{1}$ & $\mu_{2}$ & $\sigma_{2}$ & D  \\
          & [mas yr$^{-1}$] &  [mas yr$^{-1}$] &  [mas yr$^{-1}$] &  [mas yr$^{-1}$]&   \\
\cline{2-6}
Field stars & $-4.61\pm 0.11$ & $4.4 \pm 0.2$ &$-3.53\pm0.57$ &$13.63 \pm 0.85$  & $0.11\pm 0.06$\\
Position sel. & $ -4.51\pm 0.11 $&$3.52\pm0.21$ &$-4.09\pm 0.41$ &$7.05\pm 0.54$  &$0.08\pm 0.07$\\
PM selection&$-6.13\pm 0.39$ &$1.35\pm 0.16$ & $-3.52 \pm 0.41$ &$0.8 \pm 0.2$& $2.35\pm 0.13$\\
\hline\hline
\end{tabular}
\label{GMM_dec}
\end{table}

\begin{table}
\onecolumn
\centering 
\caption{Detected RR Lyrae stars within $12'$ from the Patchick 99 centre. The symbol $\bigstar$ indicates the star members of the cluster, while the double symbols $\bigstar \bigstar$ highlight the variable stars that have PM matching the mean GC PM within the errors.}
\begin{sideways}
\begin{tabular}{lccccccccccc}
\hline\hline
SourceID 	 	& RA  & DEC  &     P    &     V  &   $ K_s $ &   J  &  $D_{RRL}$   &   $d_{centre} $ & Type & $\mu_{\alpha}$ & $\mu_{\delta}$\\
   & (J2000) & (J2000)&[day]&[mag]&[mag]&[mag] &[kpc] &[arcmin]& & [mas yr$^-1$]  & [mas yr$^-1$]\\
\hline
OGLE-BLG-RRLYR-35312$^{\bigstar \bigstar}$ & 273.8098 &-29.8325& 0.4864334& 15.47 &13.84 & 14.10 &  6.1 & 8.2 & RRab & -3.866 & -3.070\\
OGLE-BLG-RRLYR-35348 &273.8499 & -29.9502 &0.2734252 &16.91 &15.57 &15.93  & 12.2 & 10.1& RRc &-0.453&-4.932\\
OGLE-BLG-RRLYR-35447$^{\bigstar \bigstar}$ & 273.9633 &-29.9070 &0.2879753 &15.93 &14.58 &14.82 & 7.9 & 5.8& RRc &6.448&-14.647\\
OGLE-BLG-RRLYR-35476$^{\bigstar}$ &274.0053 &-29.9212 &0.5301811 &15.48 &13.77 &14.16 & 6.1 & 7.4& RRab&-4.627&-4.028\\
OGLE-BLG-RRLYR-35557 &274.1133 &-29.8866 &0.6141217 &16.09 &14.26 &14.64 & 8.3 & 11.0& RRab&0.210 &-10.364\\
OGLE-BLG-RRLYR-15997 &273.7959 &-29.7434 &0.2933517 &16.18 &14.79 &15.13 & 8.7 & 9.9& RRc&-3.290&-9.259\\
OGLE-BLG-RRLYR-16094$^{\bigstar}$ &273.9599 &-29.7208 &0.5121471 &15.88 &14.27 &14.66 & 7.6 & 5.6& RRab&4.453&-3.489\\
OGLE-BLG-RRLYR-16108$^{\bigstar}$ &273.9885 &-29.7488 &0.5217077 &15.90 &14.25 &14.75 & 7.6 & 4.6& RRab&1.692&1.172\\
OGLE-BLG-RRLYR-16115 &274.0016 &-29.6816 &0.3376222 & -       &14.83 &15.08 & 9.4 & 8.6& RRc&-4.906&-8.455\\
OGLE-BLG-RRLYR-35292 &273.7955 &-29.8620 &0.2931980 &16.64 &15.06 &15.37 & 9.8 & 9.5& RRc&-3.943&-4.551\\
OGLE-BLG-RRLYR-35355 &273.8588 &-29.9444 &0.5942311 &19.28 &17.05 &17.53 & 29.4 & 9.5& RRab& -&-\\
OGLE-BLG-RRLYR-35364 &273.8741 &-29.8688 &0.5927029 &18.35 &16.51 &16.85 & 22.9 & 5.5& RRab &-4.234&-3.179\\
OGLE-BLG-RRLYR-35381 &273.8937 &-29.8050 &0.6040448 &16.61 &14.76 &15.22 & 10.3 & 3.2& RRab&-1.456&-0.187\\
OGLE-BLG-RRLYR-35382$^{\bigstar}$ &273.8947 &-29.8940 &0.6214080 &15.72 &13.91 &14.50 & 7.1 & 5.8& RRab& 2.041&1.390\\
OGLE-BLG-RRLYR-35384 &273.8970 &-29.7807 &0.7768699 &16.00 &14.17 &14.67 & 8.9 & 3.5& RRab&-3.910&-5.198\\
OGLE-BLG-RRLYR-35390 &273.9015 &-29.6619 &0.2519186 &15.26 &13.98 &14.25 & 5.7 & 9.4& RRc&-5.090&3.183\\
OGLE-BLG-RRLYR-35431 &273.9381 &-29.9478 &0.6426197 &15.90 &14.17 &14.60 & 8.1 &8.1& RRab&0.943&-1.826\\
OGLE-BLG-RRLYR-35459$^{\bigstar \bigstar}$  &273.9749 &-29.8401& 0.3754443 &15.50 &13.98 &14.30 & 6.6 & 2.4& RRc&-3.172&-8.688\\
OGLE-BLG-RRLYR-35533 &274.0821 &-29.7864& 0.3104254 &16.34 &14.96 &15.25 & 9.6 & 8.3& RRc&-1.772&-4.132\\
OGLE-BLG-RRLYR-35535 &274.0874 &-29.8810 &0.4767085 &16.07 &14.37 &14.77 & 7.6 & 9.4& RRab&-5.984&-2.042\\
OGLE-BLG-RRLYR-35571 &274.1270 &-29.8545 &0.5653381 &16.34 &14.52 &14.83 & 9.0 & 11.2& RRab&-2.534&-3.027\\
\hline\hline
\end{tabular}
\label{rrlpat}
\end{sideways}
\end{table}

\end{document}